\documentclass[12pt,preprint]{aastex}

\slugcomment{{\em Accepted by Astrophysical Journal}}

\shorttitle{ULYSSES IN-SITU INTERSTELLAR DUST FLUCTUATIONS}
\shortauthors{Strub et al.}

\begin{document}

\title{\em 
16 years of Ulysses Interstellar Dust Measurements in the Solar System: II. Fluctuations in the Dust Flow from the Data}

\author{Peter Strub~and Harald~Kr\"uger}
\affil{Max-Planck-Institut f\"ur Sonnensystemforschung, Justus-von-Liebig-Weg 3, 37077 G\"ottingen, Germany}
\email{P.Strub@gmail.com, krueger@mps.mpg.de}

\author{Veerle~J.~Sterken\altaffilmark{1}} 
 \affil{International Space Science Institute, Hallerstrasse 6, 3012 Bern, Switzerland}
\altaffiltext{1}{Max-Planck-Institut f\"ur Kernphysik, Saupfercheckweg~1, 69117 Heidelberg, Germany}

\begin{abstract}
The Ulysses spacecraft provided the first opportunity to identify and study Interstellar Dust (ISD) in-situ in the Solar System between 1992 and 2007. Here we present the first comprehensive analysis of the ISD component in the entire Ulysses dust data set. We analysed several parameters of the ISD flow in a 
time-resolved fashion: flux, flow direction, mass index, and flow width. The general picture is in agreement with a time-dependent focussing/defocussing of the charged dust particles due to long-term variations of the solar magnetic field throughout a solar magnetic cycle of 22 years. In addition, we confirm a shift in dust direction of $50^{\circ} \pm 7^{\circ}$ in 2005, along with a steep, size-dependent increase in flux by a factor of 4 within 8 months. To date, this is difficult to interpret and has to be examined in more detail by new dynamical simulations.
This work is part of a series of three papers. This paper concentrates on the time-dependent flux and direction of the ISD. In a companion paper \citep{krueger2015} we analyse the overall mass distribution of the ISD measured by Ulysses, and a third paper discusses the results of modelling the flow of the ISD as seen by Ulysses \citep{sterken2015}.
\end{abstract}

\keywords{interstellar dust, interstellar medium}

\section{Introduction}

\label{sec_introduction}

Interstellar dust (ISD) plays an important role in the formation of stars and stellar systems, the evolution of galaxies, the cooling of the interstellar medium, and in forming certain molecules in interstellar space like molecular hydrogen  ($\mathrm{H_2}$). 
The knowledge on ISD comes mainly from astronomical observations: The wavelength dependence of the extinction of starlight and hydrogen column densities provide us with information about possible ISD materials and size distributions. Observations of the interstellar gas, in comparison with  cosmic abundances of the elements (with reference to the Sun), lead to an upper limit for the mass contained in  ISD 
\citep{mathis1977,kimura2003a,slavin2008}.

The ISD mass is approximately 1\% of the total mass of the interstellar medium, and the grains are typically between nanometers and micrometers in size. They are believed to consist mainly of silicate and carbonaceous material, absorb and scatter electromagnetic radiation in the visible and in the UV wavelengths, and emit mainly in the (far) infrared \citep{draine2011}.

The Sun and the heliosphere are surrounded by a cloud of warm dense gas and dust, called the ''Local Interstellar Cloud``. Nearby is the G-cloud which the solar system will traverse within a few thousand years
\citep{frisch2011}. Because of the relative motion of the cloud and the heliosphere, interstellar gas and dust enter the solar system at a typical speed of $26.08\,\mathrm{km\,s^{-1}}$ \citep{lallement2014,wood2015a} along the interstellar gas stream direction. For some time there was an ongoing discussion about the speed, whether it is actually 23.2 or $\mathrm{26\,km\,s^{-1}}$\citep{mccomas2012}. It has now been settled to be $\mathrm{26.08\pm0.21\,km\,s^{-1}}$\citep{lallement2014,wood2015a}.

Contemporary ISD was detected in-situ in the solar system for the first time in 1992 by the Ulysses spacecraft \citep{gruen1993a}, and later confirmed by the Galileo and Helios spaceprobes, using impact-ionization detectors. The ''Stardust Interstellar Preliminary Examination'' (ISPE) team has extracted and analyzed three interstellar dust particle candidates that were captured with the Interstellar Dust Collector \citep{westphal2014b}, and interstellar dust impacts were also  identified with the Cassini Cosmic Dust Analyzer \citep{altobelli2003,altobelli2015a}. 
Finally, claims were made for detections of ISD using the  antenna system of the STEREO spacecraft \citep{belheouane2012}, and also ISD meteors were reported \citep{baggaley1999,baggaley2002}.

The Ulysses mission was particularly well suited for the detection of ISD particles. First, its almost polar orbit around the Sun with an aphelion at Jupiter's orbit (5.5 AU) took the spacecraft far above the ecliptic plane.
Given that the concentration of interplanetary dust particles (IDPs) drops at increasing ecliptic latitudes and that most of the interplanetary dust moves on prograde heliocentric orbits (Fig.~\ref{fig_3}), the orbital sections where Ulysses was neither close to its perihelion nor the solar poles were best suited for the detection of ISD. The orientation of Ulysses' orbital ellipse was such that in these sections the IDP and the ISD impact directions were almost antiparallel, and thus these populations were easy to separate.

The flow of the ISD particles in the solar system is governed by two fundamental effects: (1) the combined gravitational force and radiation pressure force of the Sun, and (2) the grain  interaction with the solar magnetic field that is ''frozen`` into the solar wind. The former effect can be described as a multiplication of the gravitational force by a constant factor $(1 - \beta$), where the radiation pressure factor $\beta = |{\bf F}_{rad}|/|{\bf F}_{grav}|$ is a function of particle composition and size. Interstellar particles approach the Sun  on hyperbolic trajectories, leading to either a radially symmetric focussing ($\beta  < 1$) or defocussing ($\beta > 1$) downstream of the Sun that is constant in time \citep{gustafson1996,landgraf2000b,sterken2013a}. Particle sizes observed by the Ulysses dust detector typically range from about 0.1\,$\mu$m to several micrometers, corresponding to $0.2 \leq \beta \leq  1.6$ \citep{gustafson1994,kimura1999}. For a detailed description of the forces acting on the particles and the resulting general ISD flow characteristics, see \citet{sterken2012a}.

The interplanetary magnetic field (IMF) shows systematic variations with time. These variations include the 25-day solar rotation (which -- to a first approximation -- averages out for ISD far from the Sun), the 22-year solar magnetic cycle, and local deviations due to disturbances in the interplanetary magnetic field. The dust particles are typically charged to a constant equilibrium potential of +5 V \citep{kempf2004}. Small particles have a higher charge-to-mass ratio, Q/m, so that their dynamics is more sensitive to the interplanetary magnetic field. The resulting dominant effect of the magnetic field on the charged dust particles is a focussing and defocussing relative to the solar equatorial plane  with the 22-year magnetic cycle of the Sun \citep{landgraf2000b,sterken2012a,sterken2013a}.

An in-depth analysis of the ISD flow with the Ulysses dust data set from 1992 to 2002 was presented by \citet{landgraf2000b} and \citet{landgraf2000a,landgraf2003}. The authors analyse the variations of the  measured ISD flux and conclude that the ISD flow is dominated by particles of radius $r_d \approx \mathrm{0.3\,\mu m}$ (corresponding to $Q/m = \mathrm{0.59\ C\,kg^{-1}}$). Finally, \citet{krueger2007b} find a shift in the direction of the dust flow  in 2005 but provide no further detailed analysis.

This paper is part of a series of three publications dedicated to the analysis of the full Ulysses data set of 16 years of interstellar dust measurements in the heliosphere. 
This work focusses on the analysis of the flux, direction, and time-dependent mass index of the ISD flow, in contrast to previous works that analysed only the flux. The final years of the Ulysses mission not only increased the inherently low number of particle detections considerably, but also reveal significant changes in directionality and mass index of the dust flow. In a  companion paper  \citep{krueger2015}, we analyse the  mass distribution of the interstellar grains measured in the heliosphere, and  a third paper presents
the results from modelling the grain dynamics \citep{sterken2015}.

\section{Dust Data}

The Ulysses spacecraft was equipped with an impact ionization dust detector that in-situ detected contemporary ISD in the solar system for the first time \citep{gruen1993a}. In this section we briefly describe the dust instrument, the data recorded by the instrument, we give an overview of the dust data set, and describe the interstellar dust component.

\subsection{The Ulysses Dust Instrument}

\label{sec_instrument}

The Ulysses dust instrument  measured the plasma resulting from hypervelocity dust impacts onto the detector target \citep{gruen1992b}. Typical impact speeds were $\mathrm{2\,km\,s^{-1}} < v_{\rm imp} < \mathrm{70\,km\,s^{-1}}$ (calibrated range). From the measured charge signals the particle's mass and impact velocity could be derived \citep{gruen1995a}. The Ulysses detector was a twin of the dust instrument flown on the Galileo mission \citep{gruen1992a}. Here we provide an overview of the instrument.

Impact ionization detectors like the one on board Ulysses rely on the fact that impacts of micrometer-sized particles onto solid surfaces at velocities above approximately $\mathrm{1\,km\,s^{-1}}$ produce a cloud of impact plasma that can be separated by an electric field into positive ions and electrons, and subsequently measured by up to three time-resolved charge detectors. Charge rise times and the total amplitude of the measured charge signal allow the determination of impact velocities and particle masses. The time correlations of independent charge detections (three for the Ulysses dust detector) allow for a suppression of noise events and, therefore, a reliable identification of real dust impacts.

The Ulysses dust detector consists of a circular entrance aperture of $\mathrm{0.1\,m^{2}}$ covered by three grids, and a spherical impact target, all at an electric potential of $\mathrm{0\,V}$. In the centre of the entrance window there is an ion collector at a potential of $\mathrm{-350\,V}$. When a particle impacts the target, the positive ions migrate towards the ion collector, and the
electrons move to the target. Both charges are measured by highly sensitive charge amplifiers. A fraction of the ions passes through the ion collector and is detected by a channeltron. All three output signals are digitized and processed to measure the signal's integrated charge and rise times of both the electron and ion channels.

The detector geometry leads to a total opening angle of $70^{\circ}$ (the angle from the instrument axis to the edge of the sensitivity zone). This is the maximum angle at which an impacting particle can hit the target. However, \citet{altobelli2004a} suggested that the side wall of the detector also contributes to the sensitive area, leading to an opening angle of $90^{\circ}$. These angles correspond to Gaussian widths of $\sigma = 25^{\circ}$ and $ 32^{\circ}$, respectively.


The Ulysses spacecraft was spin-stabilized, rotating about the axis of the high-gain antenna with a spin rate of approximately five revolutions per minute \citep{wenzel1992}. Hence, the impact direction of a dust grain is given by the sensor orientation at the time of a dust impact, which is 
defined by the spacecraft rotation angle, $\phi$.
In the time period considered in this paper, the spin axis  normally pointed towards the Earth within $1^{\circ}$. 

The dust detector is mounted at an angle of $85^{\circ}$ between its sensor axis and the spacecraft rotation axis. For most of the mission time, this led to a circular scanning pattern more or less orthogonal to the ecliptic plane, passing close to the ecliptic poles (cf. Fig.~\ref{fig_3}). A rotation angle of zero designates the point nearest to the ecliptic North direction
and $270^{\circ}$ points towards the direction of prograde heliocentric motion. In the inner solar system, close to Ulysses' perihelia, the scanning pattern was different; here the sensor scanned more or less parallel to the ecliptic plane (cf. Fig.~\ref{fig_3}).

In order to relate the measured impact charges and rise times to mass and impact velocity of the grain, the instrument's response curves were calibrated at the Heidelberg dust accelerator facility \citep{gruen1995a}. This led to a tabulated, piecewise power-law function that has been used for the data reduction. It can  be approximated by the following relation between impacting dust particle mass, $m_d$, and velocity, $v_d$, to the generated positive ion charge $Q_I$:
\begin{equation}
Q_I \sim m_d\, v_d^{3.5}. \label{equ_1}
\end{equation}

This function will be used for order of magnitude estimates in the following discussions. All measured data are digitized (6-bit integer values from 0 to 63 for the charges, and 4-bit integer values from 0 to 15 for the rise times), and stored as data records for each impact.

The typical uncertainties of the impact charges are within a factor of 1.5 for the ion charge $Q_I$, a factor of 1.7 for the electron charge $Q_E$, and about a factor of 2 for the velocity $v_d$ \citep{goeller1988}. The factor of two uncertainty in the dust velocity in combination with the
exponent of 3.5 in Equation~\ref{equ_1} leads to a large uncertainty for the mass of a factor of about 10. However, 
by assuming an inflow velocity of the ISD particles which is identical with the inflow velocity of the neutral helium gas into the heliosphere, the statistical errors can be reduced. This approach was first applied  by \citet{landgraf1998a} who assumed an inflow velocity of $\mathrm{26\,km\,s^{-1}}$ based on Ulysses measurements \citep{witte1996,witte2004b}. Here, we follow a similar approach but use a helium inflow velocity of $\mathrm{23.2\,km\,s^{-1}}$ as recently derived from IBEX measurements \citep{mccomas2012}, although this speed measurement is still under discussion \citep{lallement2014}. Recent results indicate that the original Ulysses value of $v_\mathrm{Helium}=26\,km/s$ is a better estimate for the velocity of the neutral gas \citep{wood2015a}. However, as we discuss in the following section, the effect on our analysis is very small and we therefore decided to keep using $23.2$km/s.
In so doing, we ignore variations of the flow velocity due to grain interaction with the interplanetary magnetic field. However, these variations are expected to be smaller than  30\% for dust particles with radius $r_d = \mathrm{0.2\,\mu m}$ \citep{sterken2012a}, and therefore much smaller than the uncertainties of velocities derived from the instrument calibration.

At some places in this paper we indicate particle radii. They are calculated from the measured masses assuming a spherical particle shape and a density typical of astronomical silicates $\mathrm{\rho_d \equiv \rho_{ast.sil} = 3.3\,g\,cm^{-3}}$ \citep{kimura1999}. The grain radius is  given by
\begin{equation}
r_d = \sqrt[3]{\frac{3m_d}{4\pi \rho_d}}, \label{equ_2}
\end{equation}
where $m_d$ is the dust particle mass derived from the instrument calibration. A conversion between impact charge $Q_I$, particle mass $m_d$ and particle radius $r_d$ is given in Table~\ref{tab_conversion}.

\subsection{Overview of the data set}

\label{sec_overview}

The Ulysses dust data set from the entire mission consists of impact data of 6719 dust particles \citep[upper panel of Figure~\ref{fig_1};][]{krueger2010b}. Noise events were excluded from the data set based on coincidences of the individual charge signals: for a real dust impact, at least two out of three charge measurements had to occur within a short time window \citep{baguhl1993a}.

The particles measured by the Ulysses dust detector originate from different dust populations inside and outside the solar system. The discrimination of the populations is based on data supplied by the detector and the spacecraft: impact direction, velocity, impact charge, impact time, and orbital position of the spacecraft. The different dust populations were described by \citet{landgraf1998a} and in the incremental publications of Ulysses dust data by \citet{gruen1995c} and \citet{krueger1999b,krueger2001b,krueger2006b,krueger2010b}. Here we describe the criteria used to distinguish the dust populations and discuss potential problems in the identification of ISD particles.

In order to identify ISD particles we make use of the specific geometry of the Ulysses orbit. The spacecraft's polar orbit took it up to 2.6~AU above and below the ecliptic plane, thus out of the dense parts of the zodiacal dust cloud, where interplanetary dust particles move around the Sun mostly on prograde heliocentric orbits. The orientation of Ulysses' orbital plane (i.e. the longitude of the ascending node) was such that close to the aphelion interplanetary particles on prograde orbits were detected almost opposite to the nominal flow direction of the ISD (Figure~\ref{fig_3}).

Close to Ulysses' perihelion, particles on prograde orbits approached from the same direction as the ISD. In this part of the orbit it was not possible to distinguish ISD by their impact direction from interplanetary particles on bound orbits. A clear distinction based on particle velocity is also not possible due to the large uncertainties in the velocity measurements. Therefore detections close to Ulysses's perihelion passage were excluded from the ISD data set (Tab.~\ref{tab_1}).

Four different dust populations were identified in the Ulysses dust data set. The ISD population is characterized by their impact direction and velocity being similar to the flow of the neutral interstellar gas through the solar system \citep{witte1996,witte2004a,mccomas2012,lallement2014,wood2015a}. The latter approaches from a mean upstream direction of ecliptic longitude $l = 255.7^{\circ}$ and latitude $b = 5.5^{\circ}$. We assume the average dust velocity to be $v_d = \mathrm{23.2\,km\,s^{-1}}$ as given for the flow of neutral Helium through the Solar System by \citet{mccomas2012}.
However, these values were put into question \citep{lallement2014} and the debate was recently settled with $v_{Helium} = \mathrm{26.08\pm0.21\,km\,s^{-1}}$ being the accepted value\citep{wood2015a}. We decided not to change the velocity and repeat our analysis using $v_{Helium} = \mathrm{26\,km\,s^{-1}}$ as the only effect of this would be the exclusion of 5 dust particles from the ISD dataset, reducing the overall flux by less than $0.9\%$. The mass calibration, which would be affected by the difference in velocity, has not been used in this work except for giving reference values. The longitude $l$  increases slowly by about $2.9^{\circ}$ over the total Ulysses measurement period \citep{frisch2013b}, which is also still under discussion \citep{lallement2014}. We use a value for the year 2000 as the reference direction. However, as we show in Section~\ref{sec_direction}, a shift of up to $50^{\circ}$ from the gas flow occurs in certain periods.

The remaining three populations are zodiacal dust, Jupiter dust streams, and $\beta$-meteoroids:
\begin{itemize}
\item[1)]
 The zodiacal dust population is concentrated towards the ecliptic plane and has its highest density in the inner solar system \citep{gruen1997a}. The zodiacal dust particles generally orbit the Sun on bound prograde orbits, which allows a clear separation from the ISD flow: For most of the Ulysses orbit ISD is seen from the retrograde direction, excluding only the periods around the perihelion passages. Therefore, the zodiacal dust can be removed from the data set by cutting out time intervals of 0.9~yr around the perihelion passages from a total 6.2~yr per orbit. Additional filtering of zodiacal particles is not necessary.
\item[2)]
Jupiter is an intense source of nanometer-sized dust particles concentrated in spatially and temporally constrained dust streams. The grains are ejected at very high velocities of around $\mathrm{200\,km\,s^{-1}}$ \citep{gruen1993a,zook1996}. They are only detected in short time intervals of a few days \citep{baguhl1993a,krueger2006c}. Due to their very small size these particles generate only small impact charges $Q_I \leq 10^{-13}\,\mathrm{C}$. All particles with impact charges below this value were removed from the data set.
\item[3)]
$\beta$-meteoroids are small particles that are driven radially outwards from the Solar System by the Sun's radiation pressure \citep{landgraf1998a,wehry1999}. They are measurable only in the inner solar system. These particles have already been excluded from the ISD data set by the $Q_I$-filtering  applied to suppress Jovian stream particles (see item 2 above).
\end{itemize}

These considerations led to the following filtering criteria in the data set. Detections were excluded for ecliptic latitudes $|b| < 60^{\circ}$ around Ulysses' perihelion because there ISD particles cannot be distinguished from zodiacal dust based on direction only. Additionally, intervals of identified dust streams from Jupiter were removed and spurious stream particles were excluded by filtering out impact ion charges $Q_I \leq 10^{-13}\,\mathrm{C}$.

Furthermore, detections with velocities incompatible with the ISD velocity of $\mathrm{23.2~km\,s^{-1}}$ were removed: Given that the detector's velocity measurements have a factor of two uncertainty, we ignored particles at a measured  velocity below $v_{\rm imp} = 11.6\, \mathrm{km\,s^{-1}}$. Using the newly accepted value of $v_{Helium} = \mathrm{26.08\pm0.21\,km\,s^{-1}}$ \citep{wood2015a} would barely affect our analysis: the resulting new cutoff velocity of $v_{\rm imp} = 13.04\, \mathrm{km\,s^{-1}}$ would lead to the exclusion of 5 particles from our final ISD dataset, changing the overall flux by less than $0.9\%$.

Our selection criteria for ISD grains are summarised in Table~\ref{tab_1}. The resulting data set contains 580 dust particles and is shown in the bottom panel of Figure~\ref{fig_1}.

\subsection{Data Preparation}

\label{sec_preparation}

In order to extract information about time, location, and mass dependence, the ISD data set had to be further binned into discrete time intervals and, additionally, separated by mass. However, the relatively low number of counts strictly limits the resolution in time and mass.

The time dependence of the dust properties is assessed by binning the particles in four-month intervals. In comparison with other interval sizes, this choice is the best compromise between time resolution and statistical uncertainty. This results in 2 to 37 particles with an average of 13 particles per time bin.

One important observable is the size-dependence of the ISD flow. Due to the low number of counts it was not possible to extract a statistically significant size distribution for subsets of the data set during selected time intervals. Instead, we aimed at selecting two equally sized subsets, containing the smaller and the larger portion of the ISD particles. The mass is derived from the impact ion charge $Q_I$, applying the instrument calibration. Given that the measured charge is digitised by the detector electronics into discrete charge bins, 
the data set cannot be split into exact halves. Using this method, we split the 580 ISD particles into subsets of 273 small particles ($1 \times 10^{-13}\,\mathrm{C} < Q_I \leq  8.54 \times 10^{-13}\,\mathrm{C}$) and 307 large particles ($Q_I > 8.54 \times 10^{-13}\,\mathrm{C}$). The split point corresponds to an ion impact amplitude in digital units $\mathrm{IA_{split} = 14}$ \citep{gruen1995a} and a particle radius $r_d \approx 0.24\,\mathrm{\mu m}$. The separation into small and large particles is shown in Figure~\ref{fig_2},  and approximate particle masses and sizes are given in Table~\ref{tab_conversion}.


\section{Data Analysis}

Our analysis focusses on the observed flux, the impact direction, the mass of the dust particles, and the dependence of these parameters on time and orbital position. An analysis of the ISD flux and mass during the first years of the Ulysses mission was given by \citet{landgraf2003}. We also analyze and discuss the angular width of the ISD stream and its flow direction.

\subsection{Flux}

In this section we analyse the ISD flux over the entire dust measuring period. The variability of the ISD flux during the first years of the mission was described and modelled by \citet{landgraf2003}. At the time, only dust data until 2002 was available. Here we present an analysis of the dust flux over the entire Ulysses mission extending until 2007, which exhibits features that cannot be explained by existing models \citep{landgraf2003,sterken2013a,sterken2015}.

In oder to derive the dust flux, the registered dust impacts were filtered as described in Section~\ref{sec_overview} and binned in time intervals of 0.33\,yr. In each time bin the average sensitive area $\bar{A}$ of the dust detector for particles approaching from the interstellar dust direction as well as the total measuring time for that direction (corrected for excluded time intervals) were calculated, and then used to derive the dust flux. Statistical uncertainties were propagated from the Poisson estimates for small number counts. Confidence limits were taken from \citet{gehrels1986}.

During the period from 1992 to 1995 the flux was relatively constant and somewhat higher than average. Between 1996 and 2000 it dropped to a relatively constant low level of $\mathrm{4\times 10^{-5}\,m^{-2}\,s^{-1}}$, rising again to $\mathrm{1 \times 10^{-4}\, m^{-2}\,s^{-1}}$ in 2002. In 2003 and 2004 the flux dropped to $\mathrm{7 \times 10^{-5}\, m^{-2}\,s^{-1}}$, but then increased by a factor of 3 within less than one year to reach its mission high of $\mathrm{17 \times 10^{-5}\,m^{-1}\,s^{-1}}$ in 2005. Within the following year the flux returned to its previously low value. As can be seen from the lower panel in Figure~\ref{fig_4}, the flux of the small particles reached its maximum four months earlier than that of the large particles.

The resulting dust flux shows an overall time variability by a factor of about 4 during the Ulysses mission (Figure~\ref{fig_4}, upper panel), with an average flux of $\mathrm{7 \times 10^{-5}\,m^{-2}\,s^{-1}}$. The minimum flux occurred in the interval from 1997 to 2000. The variable solar magnetic field leads to a defocussing (from 1990 to 2000) and focussing (2000 to 2011) configuration for interstellar dust due to the varying Lorentz force during different parts of the solar cycle \citep{landgraf2003,sterken2012a,sterken2013a}.  The dust spatial density in the heliosphere is determined by the time-integrated effect of the Lorentz force, and therefore lags behind the IMF configuration. In agreement with the expectation, the minimum in dust flux between 1997 and 2000 corresponds to approximately 3 years of phase lag behind the magnetic field's defocussing configuration from 1990 to 2000. This is discussed in detail in our companion paper by \citet{sterken2015}.

The rapid flux increase observed in 2005 coincides with a shift in impact direction described in Section~\ref{sec_direction}.
 At the same time the mass distribution exhibited a rapid variation due to a change in the flux of small particles ($r \lesssim \mathrm{0.24\,\mu m}$;  the separation into small and large particles is defined in Section~\ref{sec_preparation}). The small grains reach their maximum flux about 0.33 yr (1 time bin) earlier than the large particles ($r \gtrsim 0.24\,\mathrm{\mu m}$; cf. Figure~\ref{fig_4}, bottom panel). Note that a shift in impact direction observed in this period is virtually independent of the particle size within the statistical uncertainties.
 
In Figure~\ref{fig_5} we compare the flux along the three orbits of Ulysses. Given that the three orbits had practically the same detection geometry for interstellar dust, we can compare the measured fluxes at the same spatial locations but at different time intervals. 

During the inbound legs (first halfs) of orbit 1 (1992 to 1998) and orbit 3 (2004 to 2010) the flux was relatively high. The flux decreased in the second half of orbit 1 (outbound leg), whereas no dust measurements were performed in the second half of orbit 3. In the inbound leg of orbit 2 (1998 to 2004) the flux was about a factor of 4 lower than along the inbound legs of the other orbits, but increased along the outbound leg of this orbit to become about twice the flux observed in the outbound leg of orbit 1.

The highest fluxes coincide with both the focussing periods of the solar magnetic field and with the relative proximity to Jupiter. This leads to the question whether the flux variations are related to the Jupiter dust streams. While this cannot be ruled out entirely, the distribution of impact charges of the Jupiter dust streams is highly concentrated towards low impact charges $Q_I \leq \mathrm{1 \times 10^{-13}\, C}$. In the time intervals with increased ISD flux, the small particles contribute more strongly to the total flux, but there is no indication for an increase of the very small particles with $Q_I \leq \mathrm{1 \times 10^{-13}\, C}$ which would indicate contamination from Jupiter stream particles outside the  time intervals when dust streams were identified. If the flux increase were indeed related to the proximity to Jupiter, this would likely be a population of particles with properties distinct from both the average ISD population and the known Jupiter streams so that they would remain unrecognised by our particle
selection criteria.

\subsection{Methods: Fitting with Low Number Statistics}

\label{sec_methods}

In order to determine particle properties like the two-dimensional direction or the angular width of the ISD flow, we fitted a model for
the angular distribution of the interstellar dust to the measured data. Most properties of the dust data discussed in this work are derived from small number count data that follow Poisson statistics. In this low-count regime, $\chi^2$-statistics leads to significant systematic fitting errors for sample sizes below 100 and is no longer a valid approximation. In our data analysis, data bins often contain less than 10 particles. Therefore a maximum likelihood approach is required for an accurate determination of the quality of fit and the confidence levels.

We follow the approach laid out by \citet{cash1979}. In this method, the $\chi^2$ as a measure for the goodness of fit is replaced by the Cash statistics function, or C-statistics, which is then minimized by varying the parameters of the model function in a standard fitting algorithm:
\begin{equation}
C = 2\sum_{i=1}^{N} (e_i - n_i \ln e_i) \equiv 2 \left (E-\sum_{i=1}^{N} n_i \ln e_i \right ) \label{equ_3}
\end{equation}
This assumes a data set of Poisson distributed data in $N$ data bins with $n_i$ counts in a bin $i$, the corresponding model values (expected counts) $e_i$ in the same bins, and the sum of the model counts $ E = \sum e_i$.

For large count numbers, the cash statistics function converges toward the $\chi ^2$ statistics, but remains accurate down to 1 count per bin. In this formulation the location of the bins for data and model along an independent variable (i.e. x axis) is irrelevant.

The maximum-likelihood approach is used to determine the quality of fit in two cases:
\begin{itemize}
\item To determine the direction of the dust flow on the celestial sphere. Each point on a mesh of ecliptic latitudes and longitudes (in steps of $1^{\circ}$) has been considered as a possible dust inflow direction. Based on the Ulysses orbit geometry, we calculated the relative dust flux as a function of the rotation angle $\phi$, integrated over the observed time interval. We assumed a uniform monodirectional flow for the incoming dust, widened by the angular sensitivity profile of the detector which is a function of the impact angle only \citep{gruen1992a}. For a given direction, the modelled sensitivity at each rotation angle is therefore determined for the angle between the detector axis and the flow direction. Then the C-statistics was determined by comparing the rotation angle data to the model distributions. This results in a 2-dimensional distribution of C-values over the celestial sphere, which were then used to plot the contours of $1\sigma$, $2\sigma$ and $3\sigma$ significance, as determined from a $\chi ^2$ distribution for zero degrees of freedom.
\item To determine the width of the dust flow. Based on the detector's angular sensitivity function \citep{gruen1992a} we assumed a broadened monodirectional flow and approximated the measured angular distribution of impacts by a Gaussian distribution. More complex models did not provide reasonable stability of fitting and other simple models gave indistinguishable differences due to the relatively low number of impacts. The Gaussian width and normalization were determined as the best fit from minimizing the C-statistics with a standard fit algorithm available in IDL 7 \citep[Amoeba fit,][]{press1992}. We derived confidence intervals by varying the best-fit parameters until the statistics reached values corresponding to $1 \sigma$, $2 \sigma$ and $3 \sigma$ significance for two degrees of freedom.
\end{itemize}

\subsection{Direction of the ISD Flow}

\label{sec_direction}

Another key property of the ISD component is the direction of the flow. The opening angle of the detector allows us to constrain the impact direction of a single particle to within $\pm 70^{\circ}$, or $\pm 90^{\circ}$ if we take the sensor side wall into account. Using statistical analysis with a sufficient number of particles, accuracies of about ten degrees can be achieved. 

Given that the impact direction of each grain is derived from the spacecraft rotation angle at the time of the particle impact, only this single variable is available to constrain the direction of the interstellar dust flow. Therefore, we first analyse the ISD flow direction in one dimension (i.e. the Ulysses rotation angle $\phi$;  Section~\ref{sec_1-dim}). In a second step, we derive the ISD flow direction in two dimensions on the celestial sphere in ecliptic coordinates (Section~\ref{sec_2-dim}). Finally, we analyse the width of the angular distribution of the interstellar dust flow (Section~\ref{sec_width}).

\subsubsection{1-Dimensional (Spacecraft Rotation Angle $\phi$, Time-Resolved)}

\label{sec_1-dim}

The circular scan pattern of the detector on the sky practically reduces the direction measurement to one dimension along which directional shifts can be observed.
In order to eliminate the influence of Ulysses' orbital motion on the scan orientation, we calculated the expected impact direction of an undisturbed flow approaching from the direction of the interstellar helium flow for all times of the mission. We then calculated the average approach direction in all time intervals relative to this nominal direction. The resulting angular deflection $\Delta \phi$ can be seen in Figure~\ref{fig_6} for all ISD particles and in Figure~\ref{fig_7} separately for small and large particles as defined in Section~\ref{sec_preparation}.

We found a couple of remarkable results. During some periods strong deviations occurred from the nominal flow direction that exceed $50^{\circ}$ (1999, 2005). However, when taking the statistical uncertainties into account, the most striking feature is the shift of the ISD direction in 2005 which turned out to be highly significant at a level of more than $ 3 \sigma$. In this period the deviation from the assumed inflow 
direction was $50^{\circ} \pm 7^{\circ}$ .

The data indicates more directional variability during the entire Ulysses mission. In Figure~\ref{fig_6}, for about half of the time bins, the significance of a deviation from the nominal direction exceeds  $1\,\sigma$. For each of these individual bins this is not significant (which would require $3\,\sigma$).
It indicates, however, that there is more variation than is compatible with an undisturbed monodirectional flow, which would require 68\% of the particles to be within $ 1\,\sigma$ of the nominal flow. This is more pronounced for the small particles  ($r \lesssim 0.24 \mathrm{\,\mu m}$),
 whereas the large particles are largely compatible with an undisturbed statistical distribution (see also Section~\ref{sec_width}). Most of the time-binned directions deviate by less than $2\,\sigma$ from the gas flow direction, while relatively few ($\lesssim$ 30\%) are within $1\,\sigma$ of the flow. Obvious features are tendencies on a $1\,\sigma$ level towards rotation angles smaller than the helium direction during the period from January 1993 until June 1994, and to larger rotation angles in 1995 and 1996, respectively.

\subsubsection{2-Dimensional Flow Direction (Longitude, Latitude)} 

\label{sec_2-dim}

We demonstrated in the previous section that in 2005 the ISD flow showed a significant shift in rotation angle away from the Helium inflow direction. In order to constrain the flow direction on the celestial sphere in ecliptic coordinates we use the fit method described in Section~\ref{sec_methods}.

In order to determine the shift in impact direction during the period of the strongest deflection in 2005, we split the ISD data set into two time intervals: time interval 1 includes particles measured from 1992 to 2004 and in 2006, when we observed no significant deviation from the Helium direction. Time interval 2 includes specifically the year 2005, which is the period with the strong shift in impact direction. Figure~\ref{fig_8} shows the resulting  upstream directions in ecliptic coordinates. The
definition of  interval 2 was based on both the angular shift and the flux increase which occurred in all three time bins in 2005, but are either absent or not significant earlier and later.

The dust impact direction observed in time interval 1 is constrained at the ecliptic longitude $l = 255^{\circ} \pm 30^{\circ}$ and latitude $b = 13^{\circ} \pm 4^{\circ}$ and is in good agreement with the interstellar Helium flow direction measured by IBEX \citep[$l = 255.7^{\circ}$, $b = 5.4^{\circ}$,][]{frisch2013b}. Due to the scanning pattern of the dust detector that swept relatively close to the ecliptic North-South direction, the uncertainties in the determination of the dust inflow direction are much smaller in ecliptic latitude than in longitude.

In time interval 2, the uncertainties in the dust flow direction are much larger which is mostly
due to the lower detection statistics in the shorter time interval. The best-fit direction
is $l = (289^{+15}_{-60})^{\circ}$ and $b = (-38^{+20}_{-25})^{\circ}$. Uncertainties correspond to the limits of the $1\,\sigma$ contour in Figure~\ref{fig_8}. The shift 
in flow direction is highly significant well beyond the $3\,\sigma$ level. It is clearly identified in ecliptic latitude while the large uncertainties in ecliptic longitude prohibit any strong  constraints of the shift in that coordinate.

\subsubsection{Width of the angular distribution}

\label{sec_width}

For a mono-directional dust flow, the measured width of the distribution of impact angles depends only on the orbit geometry and the detector's sensitivity as a function of  impact angle. Deviations from this expected angular distribution can be used to quantify the deflection of the dust flow from a mono-directional flow. 

As in the previous sections, we describe the angular distribution of the inflowing ISD as a normalized Gaussian to account for broadening over the geometrical model of a mono-directional flow. We fitted the width of the Gaussian to the angular distribution in each time bin (Section~\ref{sec_methods}). We performed this fitting for the full data set containing all particles as well as for the subsets separated into small and large particles. The resulting width as well as its $1\,\sigma$-uncertainty are shown in Figure~\ref{fig_9}. The nominal sensor opening angle of $\pm 70^{\circ}$ corresponds to $25^{\circ}$ Gaussian width, while the opening angle including wall impacts of $\pm 90^{\circ}$ corresponds to $32^{\circ}$ Gaussian width.

As can be seen in the upper panel of Figure~\ref{fig_9}, the measured width of the angular distribution  of the ISD particles 
is typically larger than expected from the instrumental profile without wall impacts. Periods of low flux have poor statistics, as can be seen from large error bars that are also consistent with zero width. For all but two cases the width exceeds the nominal detector width by more than $1\sigma$. However, when  wall impacts are taken into account for the detector opening angle, during four time intervals the width is compatible with the detector geometry.

The small particles  exhibit larger measured widths than the big particles (Figure~\ref{fig_9}, middle and bottom panels): The large particle flow has a typical width of $34.2^{\circ} \pm 5.4^{\circ}$, whereas the small particles have a width of $44.1^{\circ} \pm 5.0^{\circ}$. Also the large particle widths deviates by more than $1\sigma$ from the detector width in only three time bins, whereas the small particles deviate by more than $2\sigma$ in six time bins.
It shows that the width derived for the small particles is significantly larger than that of the large particles.

In general, the measured widening is a combination of several effects. Firstly, a widening of the incoming dust flow itself leads to a widening of the observed flow. Secondly, as the particles used for the width measurement are integrated over intervals of 1 year, a fluctuation of the direction of a mono-directional stream would also lead to a measurable widening. These two effects cannot be distinguished because the count statistics is a limiting factor here. Both shall be considered as intrinsic widening of the ISD flow.

Finally, a combination of Ulysses' orbital geometry, the orientation of the detector scanning pattern, and a shifted dust flow direction can widen the measured width. This detection window is a combination of the detector's large opening angle of $\pm90^{\circ}$ \citep[assuming impacts onto the detector side walls,][]{altobelli2004a} and the angle of $85^{\circ}$ between the detector boresight and the rotation axis, which lead to a small projected sensor area for a dust flow aligned with the rotation axis. Such a flow would be detectable at the full $360^{\circ}$ of rotation angles. A widening by more than $10\%$ of the original width due to this effect can only be achieved for dust flowing within $25^{\circ}$ of the rotation axis. We shall refer to this as geometric widening.

Is the measured width an intrinsic widening of the ISD flow or a geometric effect? While this cannot be answered with certainty, there are hints towards an intrinsic widening. The widening is only apparent in the population of small particles, whereas the large particles are compatible with a mono-directional flow. Simulations by \citet{sterken2012a} show that this is expected if the widening is caused by the interaction with the solar magnetic field, which is effective for particle masses in our "small particles" dataset, but barely affects particle sizes representative of our "large particles" dataset.

For the small particles, time bins that exhibit a high measured width also coincide with intervals of large angular shifts and/or high variability of the direction. This is an indication that the increase in measured width may be due to increased directional variability.

{he conditions leading to a significant geometric widening that can explain the observations would require a dust inflow within $<25^{\circ}$ of the spacecraft rotation axis. During the whole Ulysses mission, the nominal flow direction of the ISD was always $>55^{\circ}$ away from the spacecraft's rotation axis. Hence, explaining the observed widening by geometric effects requires the flow direction to shift by at least $30^{\circ}$. While this cannot be excluded per se, it would mean two things: The shift has to occur both by the correct angle and direction towards the rotation axis, and also the particles would hit the detector at a grazing angle, reducing the effective sensitive area to 10-30\% compared to the nominal direction. This would mean that the actual ISD flux would have to be $3-10$
times higher than the measured flux, higher than most of the mission, which is unlikely.

This leaves us with two possible scenarios, or a mix of both. If we interpret the intervals of high measured width as an increase in flow width, this could be due to a widening of the ISD flow itself or fluctuations of the direction on timescales smaller than the bin size of one year. If we assume geometric widening to be the main cause, the direction of the ISD flow has to change by an angle of more than $30^{\circ}$ in combination with an increase in flux by a factor of 3-10. While these cases remain indistinguishable by means of the Ulysses dust detector, both scenarios require significant changes of the ISD flow direction by at least 30$^\circ$ on timescales under a year.

\subsection{Mass-Dependent Effects}

An important property of the dust flow is its grain mass distribution. Given that the forces acting on the particles (Lorentz force, gravity, radiation pressure) depend on the particle mass, temporal variations of the mass distribution can help to constrain the effect of these forces and, ultimately, the physical properties of the particles. However, the relatively low number of ISD particles in the Ulysses dust data imposes tight limits on the resolution in time and mass. Instead of considering mass-resolved mass distributions we therefore determine the mass index defined as follows:
\begin{equation}
\mathrm{MI} = (n_{\rm big} - n_{\rm small})/(n_{\rm big} + n_{\rm small}). \label{equ_4}
\end{equation}
MI measures the relative contributions of particles smaller and larger than the median mass. A bin size of 1 year is a good compromise between time resolution and statistical uncertainties.

The time dependence of the mass index is shown in Figure~\ref{fig_10}. From 1995 to 1998, the large particles clearly dominate ($\mathrm{MI > 0}$), whereas from 2003 to 2005 small particles dominate. In 1992/1993, from 1999 to 2002 and again in 2006 the fluxes of both sizes are comparable.

This may again be understood in terms of the focussing and defocussing of the solar magnetic field due to the 22-year solar cycle. In the years after 2000, the solar magnetic field changed to a focussing configuration: The smaller particles, which are more sensitive to electromagnetic forces than the larger ones due to their higher charge-to-mass ratio, are more focussed, thus leading to a lower mass index ($\mathrm{MI < 0}$).

\section{Discussion}

\label{sec_discussion}

Thus far in this paper, we have discussed the data selection, the observed flux, flow direction, variations in the angular width of the dust flow, and the relative proportion of large to small particles detected during the Ulysses mission. Here, we briefly compare these results to simulations of interstellar dust at Ulysses' orbit. The details of the simulations are described by \citet{sterken2012a,sterken2015} and allow us to better understand in a qualitative manner why such variations occur. A quantitative comparison of data and simulations is the topic of future work.

\subsection{Simulated and Observed Flow Characteristics}

The flux, flow direction and impact speed of interstellar dust as observed by Ulysses are affected by the position and the velocity of Ulysses with respect to the interstellar dust flow. Additionally, inside the heliosphere, small (sub-micrometer sized) grains are affected by their interaction with the solar wind magnetic field. This effect increases for smaller particles and leads to a strong depletion of nanometer-sized interstellar grains in the heliosphere\citep{sterken2013a}.

Dynamical simulations show that the flux of the biggest particles ($m_d > 5 \times 10^{-15}\,\mathrm{kg}$, which, according to Equation~\ref{equ_2}, corresponds to $r_d \approx 0.7\,\mathrm{\mu m}$) increases by up to a factor of 2 to 2.5 at every perihelion passage, mainly due to the variation in relative velocity between Ulysses and the interstellar dust. These rather big particles are barely affected by the Lorentz force and their flux  depends only on Ulysses' location in its orbit. The smaller the ISD particles, the more susceptible they are to the Lorentz force. Thus, the flow of the small particles in our data set is modulated by the time-variable Interplanetary Magnetic Field (IMF) which changes periodically with the solar cycle. As a result, the simulated flux of small particles decreases between about 1996 and 2004. On the other hand, the peaks in the flux of the biggest particles around Ulysses' perihelia in 1995 and 2007 are enhanced by the Lorentz force due to the smaller particles. Finally, the peak in the flux for big particles at the perihelion in 2001 is reduced by the Lorentz force acting on the small ISD particles. The lowest flux in the simulations occurs around 1998 when a low flux is also seen in the data. As from 2000, the simulated flux increases again, also in agreement with the Ulysses measurements.

Furthermore, the changes in the dust impact direction of the biggest simulated particles depend only on the position of Ulysses along its orbit: At the perihelia, shifts of $25^{\circ}$ in longitude  and $30^{\circ}$ in latitude occur. Note that these time periods are removed from the data set because ISD particles cannot safely be separated from interplantery impactors in this part of the Ulysses orbit (cf. Table~\ref{tab_1}). For mid-sized particles (around $0.35\,\mathrm{\mu m}$) -- having higher $\beta$-values 
 -- the shifts in direction at the perihelia are smaller for $\beta \approx 1$ and $Q/m < 0.5\,\mathrm{C\,kg^{-1}}$. Larger $\beta$-values (e.g. $\beta = 1.5$, typical for particles of about $\mathrm{0.2\,\mu m}$) result in larger changes in the
direction of the dust at the perihelia of up to $80^{\circ}$ in latitude and $40^{\circ}$ in longitude and in the opposite direction for the particles with $\beta < 1$.

The data at the perihelia were excluded from the ISD data set (see Section~\ref{sec_overview}). Thus, it is not possible to see the peaks in flux and the dramatic changes in impact direction for the big particles around the perihelia in 1995, 2001 and 2007 that are  predicted by the simulations \citep{sterken2015}. However, a few big particles ($m_d > 10^{-13}\,\mathrm{kg}$, corresponding to $r_d \approx 1.9\,\mathrm{\mu m}$ according to Equation~\ref{equ_2}) were indeed detected mainly around Ulysses' perihelia. This may correspond to an increase in the flux of big ISD particles, however,  it could also have other reasons (e.g. cometary dust stream particles). A deeper investigation is needed to determine the source of these grains.

The modulation of the fluxes derived for big and small particles, and especially the variation in the mass index (see Fig.~\ref{fig_10}) which is a measure of the relative contributions of large and small particles in the ISD flow, indicates that the ISD flow is indeed modulated by the solar cycle. This is consistent with simulations by \citet{sterken2015} and was already seen by \citet{landgraf2003} from the data collected between 1992 and 2002. Note that these simulations did not take into account the filtering and focussing effects at the heliopause \citep{linde2000,slavin2012}. Between the perihelion passages of Ulysses, little variability is seen for the big particles in the simulations. Small particles (with increasing charge-to-mass ratio) show more variability in impact direction and flux around the perihelia, and the smaller the particles get, this variability extends more and more to the periods before and after the perihelia. A more detailed discussion of the simulations in the context of the measurements is given by \citet{sterken2015}.

The velocity of the neutral gas flow -- and also the ISD embedded in the medium -- is subject to an ongoing debate. In this analysis we use a gas velocity of v=23.2\,km/s as given by \citet{mccomas2012}, whereas \citet{lallement2014} suggest a velocity of 26\,km/s. In this range, the particular choice of ISD velocity has little impact on our results, as it only affects the lower velocity limit of dust particles included in the dataset. If we assume an ISD velocity of 26\,km/s instead of 23.2 km/s, we obtain a slightly smaller total number of ISD particles of 575 instead of 580 (a reduction by 0.9\%), the effect of which is well below the statistical uncertainties. A dust velocity of 26\,km/s would also affect the derived particle masses, leading to a decrease in mass by a factor of $(23.2/26)^{3.5}= 0.67$ (Equation~\ref{equ_1}). This changes the measured mass distribution of the ISD, which is discussed in detail in the companion paper \citet{krueger2015}. Here we use the impact charge as a measure of the particle mass (masses are only given for reference), thus the results of this paper remain unaffected by this factor.

\subsection{Improving Data and Simulations}

A trend in the flux for small and large particles that follows the solar cycle can be identified by comparing the data to the simulations for the whole period of the Ulysses dust data set, confirming earlier results by \citet{landgraf2003}. However, some aspects (for instance, the shift of the dust direction) are not yet fully understood and, thus, a detailed comparison of the simulations and the data remains to be done. This has to be an iterative process, where simulations improve the data analysis through an improved estimation of the impact velocity \citep{landgraf2000a}, and the fit of the data to the simulations improves the simulations through better determination of the particle properties. As a result, we will also be
able to better constrain the properties of the ISD particles. The tools and the data are now available to start this process.

Several other improvements can also help to improve data analysis and simulations: new calibration experiments of impact ionization detectors for particles with lower bulk densities (porosity) are to be made \citep{sterken2012b,sterken2015}, the ISD modelling shall include a module for simulating the effect of the heliopause, and use IMF data to calculate Lorentz forces instead of using a cyclic theoretical model for the IMF.

\section{Summary and Conclusions}

In this paper we analyzed the entire data set of Interstellar Dust (ISD) as recorded by the Ulysses dust detector between 1992 and 2007. We focussed on  variations of the ISD flux, directionality, and measured width for all particle masses. We also compared these variations for two subsets of masses (''large`` and ''small`` particles) and put the results into perspective to the simulations by \citet{sterken2012a,sterken2015}. The mass distribution is studied in detail in the companion paper by \citet{krueger2015}. A detailed comparison of the flux, direction, flow width and mass distributions with dynamical modelling is subject to a follow-up project, and shall follow the recommendations from Section~\ref{sec_discussion}.

We conclude that the observations of the dust flux show a variability over the entire Ulysses data set that is correlated with the solar cycle, confirming the results of \citet{landgraf2003}. The full Ulysses ISD data set provides us with a larger total number of particles and therefore better statistics for comparison of mass distributions with simulations. Several details like the observed shift in dust direction in 2005 are not yet fully understood but they are identified  and characterized. Moreover, we now have the basic tools to start the iterative process between improving data analysis and simulations, in order to understand the more detailed features in the data. In this process, the data analysis can be improved by using the simulation results, and the simulations can be improved by fitting them to the data.

\subsection*{Acknowledgments}

We thank the Ulysses project at ESA and NASA/JPL for effective and successful mission operations. This research was supported by Deutsche Forschungsgemeinschaft (DFG) grant KR 3621/1-1. PS would like to thank Mih\'{a}ly Hor\'{a}nyi for his hospitality and the helpful discussions during a visit in Boulder, and Eberhard Gr\"{u}n for many fruitful discussions. PS thanks the MPI f\"ur Sonnensystemforschung and the University of Stuttgart for their support. HK acknowledges support by the German Bundesministerium f\"ur Bildung und Forschung through Deutsches Zentrum f\"ur Luft- und Raumfahrt e.V. (DLR, grant 50 QN 9107). 
We are grateful to  an anonymous referee for improving the presentation of our results and for valuable additions to the data analysis.

{\it Facilities:} \facility{Ulysses}.

\appendix

\section{Appendix}

The data shown in Figures~\ref{fig_4}, \ref{fig_6} and \ref{fig_7} are listed in 
Tables~\ref{tab_3} to \ref{tab_5}.

\bibliography{./paper}

\clearpage

\section*{Tables}

\begin{table}[th]
\caption{Conversion between impact charge $Q_I$, particle mass $m_d$ derived from the instrument calibration for an impact speed of $23.2\,\mathrm{km\,s^{-1}}$ \citep{gruen1995a},
 and approximate particle radius $r_d$ derived from Equation~\ref{equ_2}, assuming a particle density  $\rho_d = 3300\,\mathrm{kg\,m^{-3}}$
. If a dust velocity of 26\,km/s is assumed, the derived particle masses decrease by a factor of $(23.2/26)^{3.5}= 0.67$ (Equation~\ref{equ_1}). }
\label{tab_conversion}
\vskip4mm
\centering
\begin{tabular}{ccc}
\hline 
Impact charge             & Particle Mass$^{\dagger}$        & Particle Radius \\
$Q_I$  $\mathrm{[C]}$     &  $m_d$ [kg]                      &  $r_d$ [$\mathrm{\mu m}$]\\
\hline
$1.0  \times 10^{-13}$    &  $2.3 \times 10^{-17}$           &             0.12    \\
$8.54 \times 10^{-13}$    &  $2.0 \times 10^{-16}$           &            0.24  \\
$6.1  \times 10^{-11}$    &  $1.4 \times 10^{-14}$           &            1.0   \\
\hline
  & & \\
\end{tabular}

$\dagger$ Note that these masses do not apply to the Jupiter stream particles which are smaller and faster than implied by the dust instrument calibration \citep{zook1996}.
\end{table}

\begin{table}[th]
\caption{Filtering criteria for the Interstellar Dust population. Notice that these criteria differ from those used in the companion paper by \citet{krueger2015} because of a different focus in the analysis. 6139 out of the full data set of the 6719 particles in the full Ulysses dust data set fulfil these exclusion criteria.
\label{tab_1}}
\vskip4mm
\centering
\begin{tabular}{lll}
\hline 
Criterion                          &    Time Period/           & Comments \\
                                   &  Spatial Region           &          \\
\hline
All dust impacts excluded in       &  1992/1993 and            &  Removal of Jupiter dust streams \\
39 short time intervals defined    &  2002 to 2005             &          \\ 
by \citet{baguhl1993a} and          &                           &          \\
\citet{krueger2006c}                &                           &          \\
\hline 
$Q_I \leq 1 \times 10^{-13}\,\mathrm{C}$ ignored & Entire data set         & Removal of Jupiter dust streams \\
\hline
$v_{\rm impact} \leq 11.6\,\mathrm{km\,s^{-1}}$ ignored & Entire data set        & Exclude particles on bound orbits \\
\hline
All dust impacts ignored with      & Inner solar system        & Remove orbit section close to perihelion \\
ecl. latitude $|b| < 60^{\circ}$ on peri-  &                   & where ISD and prograde zodiacal par- \\
helion side                        &                           & ticles cannot be separated by direction. \\
                                   &                           & This effectively excludes 0.9\,yr around \\
                                   &                           & each perihelion.     \\
\hline
\end{tabular}
\end{table}

\begin{table}[h]
\caption{Flux and impact direction for the full data set of interstellar grains derived in this paper. 
 Column (1) 
lists the center of the time interval, col. (2) the flux averaged during the time interval, 
col. (3) and (4) give the lower and upper uncertainties in the flux, col. (5) lists the deviation of 
the rotation
angle from the helium flow direction $\Delta \phi$, and col. (6) lists the uncertainty in $\Delta \phi$. 
The data are shown in the top panel of Figure~\ref{fig_4} and in Figure~\ref{fig_6}.
 \label{tab_3} }
\vskip4mm
\centering
\tiny
\begin{tabular}{cccccc} \hline
  Time            &   Flux                     &  Err Flux-                    &   Err Flux+  & $\Delta \phi$  &   Err $\Delta \phi$ \\
$\mathrm{[Year]}$ &$\mathrm{[m^{-2}\,s^{-1}]}$ & $\mathrm{[m^{-2}\,s^{-1}]}$  &$\mathrm{[m^{-2}\,s^{-1}]}$ & $\mathrm{[deg]}$ & $\mathrm{[deg]}$ \\
  (1)   &    (2)              &   (3)   &    (4)    &    (5)    &    (6)   \\
 \hline
1992.17 &  1.84E-04 &  3.09E-05 &  3.65E-05 &   0 &   7\\
 1992.50 &  1.06E-04 &  2.29E-05 &  2.84E-05 &  17 &   8\\
 1992.83 &  7.95E-05 &  1.91E-05 &  2.43E-05 &  -1 &  11\\
 1993.17 &  8.75E-05 &  1.94E-05 &  2.43E-05 &  -9 &   6\\
 1993.50 &  7.34E-05 &  1.87E-05 &  2.42E-05 & -31 &  16\\
 1993.83 &  9.18E-05 &  1.90E-05 &  2.34E-05 & -17 &  10\\
 1994.17 &  5.50E-05 &  1.45E-05 &  1.90E-05 & -16 &   8\\
 1994.50 &  5.89E-05 &  1.75E-05 &  2.36E-05 & -34 &  24\\
 1994.83 &  9.27E-05 &  1.88E-05 &  2.30E-05 &  -1 &  12\\
 1995.50 &  3.01E-05 &  1.30E-05 &  2.03E-05 &  19 &  35\\
 1995.83 &  6.27E-05 &  2.31E-05 &  3.37E-05 &  21 &  12\\
 1996.17 &  6.63E-05 &  1.75E-05 &  2.29E-05 &  21 &  13\\
 1996.50 &  3.14E-05 &  1.16E-05 &  1.69E-05 &  24 &  11\\
 1996.83 &  4.32E-05 &  1.49E-05 &  2.13E-05 &  19 &   9\\
 1997.17 &  1.86E-05 &  8.89E-06 &  1.47E-05 & -12 &  34\\
 1997.50 &  3.59E-05 &  1.24E-05 &  1.77E-05 &   1 &  11\\
 1997.83 &  3.71E-05 &  1.28E-05 &  1.83E-05 &   8 &   7\\
 1998.17 &  3.61E-05 &  1.25E-05 &  1.78E-05 &   3 &   9\\
 1998.50 &  3.23E-05 &  1.19E-05 &  1.74E-05 &  -5 &  10\\
 1998.83 &  3.01E-05 &  1.11E-05 &  1.62E-05 & -14 &   7\\
 1999.17 &  3.08E-05 &  1.14E-05 &  1.66E-05 & -14 &  21\\
 1999.50 &  1.93E-05 &  9.23E-06 &  1.53E-05 &   1 &  20\\
 1999.83 &  8.11E-06 &  5.22E-06 &  1.07E-05 &  56 &  43\\
 2000.17 &  2.85E-05 &  1.05E-05 &  1.53E-05 & -15 &  17\\
 2000.50 &  3.73E-05 &  1.37E-05 &  2.00E-05 &  26 &  30\\
 2000.83 &  7.16E-05 &  1.59E-05 &  1.98E-05 & -17 &  11\\
 2001.17 &  7.42E-05 &  2.20E-05 &  2.98E-05 &   9 &  12\\
 2001.83 &  4.63E-05 &  2.21E-05 &  3.65E-05 &   2 &  23\\
 2002.17 &  5.70E-05 &  1.69E-05 &  2.28E-05 &  19 &  15\\
 2002.50 &  1.08E-04 &  2.19E-05 &  2.68E-05 &  17 &  11\\
 2002.83 &  1.03E-04 &  2.81E-05 &  3.72E-05 &  12 &  18\\
 2003.50 &  6.63E-05 &  2.62E-05 &  3.95E-05 &  19 &  29\\
 2003.83 &  7.19E-05 &  2.04E-05 &  2.73E-05 & -73 &  24\\
 2004.17 &  5.33E-05 &  1.65E-05 &  2.27E-05 & -13 &  22\\
 2004.50 &  3.78E-05 &  1.80E-05 &  2.98E-05 & -25 &  29\\
 2004.83 &  7.66E-05 &  1.89E-05 &  2.43E-05 & -25 &  12\\
 2005.17 &  9.81E-05 &  2.23E-05 &  2.81E-05 &  27 &  15\\
 2005.50 &  1.66E-04 &  3.07E-05 &  3.70E-05 &  50 &   7\\
 2005.83 &  1.51E-04 &  2.48E-05 &  2.92E-05 &  35 &   8\\
 2006.17 &  1.25E-04 &  2.26E-05 &  2.72E-05 &  16 &   5\\
 2006.50 &  6.03E-05 &  1.71E-05 &  2.29E-05 &  32 &  11\\
 2006.83 &  1.11E-05 &  6.05E-06 &  1.08E-05 &   0 &   9\\
 2007.17 &  2.00E-05 &  7.93E-06 &  1.19E-05 &  25 &  22\\
 \hline
 \end{tabular}
 \end{table}

\begin{table}[h]
\caption{Same as Table~\ref{tab_3} but for the small particles with impact charges $10^{-13}\,\mathrm{C} < Q_I \leq 8.54 \times \mathrm{10^{-13}\,C}$. The data are shown in the bottom panel of Figure~\ref{fig_4} and in the top panel of Figure~\ref{fig_7}. 
 \label{tab_4} }
\vskip4mm
\centering
\tiny
\begin{tabular}{cccccc} \hline
  Time            &   Flux                     &  Err Flux-                    &   Err Flux+  & $\Delta \phi$  &   Err $\Delta \phi$ \\
$\mathrm{[Year]}$ &$\mathrm{[m^{-2}\,s^{-1}]}$ & $\mathrm{[m^{-2}\,s^{-1}]}$  &$\mathrm{[m^{-2}\,s^{-1}]}$ & $\mathrm{[deg]}$ & $\mathrm{[deg]}$ \\
  (1)   &    (2)              &   (3)   &    (4)    &    (5)    &    (6)   \\
 \hline
 1992.17 &  1.10E-04 &  2.38E-05 &  2.96E-05 &   0 &   8 \\
 1992.50 &  4.53E-05 &  1.48E-05 &  2.07E-05 &  17 &  12 \\
 1992.83 &  5.61E-05 &  1.59E-05 &  2.13E-05 &  -6 &  12 \\
 1993.17 &  7.00E-05 &  1.73E-05 &  2.22E-05 & -10 &   6 \\
 1993.50 &  2.94E-05 &  1.16E-05 &  1.75E-05 & -43 &  20 \\
 1993.83 &  4.39E-05 &  1.30E-05 &  1.76E-05 & -36 &  14 \\
 1994.17 &  7.86E-06 &  5.06E-06 &  1.03E-05 &  -5 &  17 \\
 1994.50 &  2.68E-05 &  1.15E-05 &  1.81E-05 & -96 &  42 \\
 1994.83 &  3.86E-05 &  1.20E-05 &  1.65E-05 & -20 &  22 \\
 1995.50 &  1.20E-05 &  7.76E-06 &  1.58E-05 & 135 &  82 \\
 1996.17 &  1.42E-05 &  7.71E-06 &  1.38E-05 &  77 &  51 \\
 1996.50 &  4.49E-06 &  3.72E-06 &  1.03E-05 &  72 &  35 \\
 1996.83 &  1.62E-05 &  8.79E-06 &  1.57E-05 &  46 &   4 \\
 1997.17 &  4.65E-06 &  3.86E-06 &  1.06E-05 &  34 &  35 \\
 1997.50 &  8.98E-06 &  5.79E-06 &  1.18E-05 &  26 &  35 \\
 1997.83 &  1.39E-05 &  7.55E-06 &  1.35E-05 &  -2 &   7 \\
 1998.17 &  1.36E-05 &  7.35E-06 &  1.31E-05 &  26 &   5 \\
 1998.50 &  9.23E-06 &  5.95E-06 &  1.21E-05 &   6 &  13 \\
 1998.83 &  1.72E-05 &  8.21E-06 &  1.36E-05 & -13 &  11 \\
 1999.17 &  1.32E-05 &  7.17E-06 &  1.28E-05 & -45 &  31 \\
 1999.50 &  4.84E-06 &  4.01E-06 &  1.11E-05 &  15 &  35 \\
 1999.83 &  4.06E-06 &  3.36E-06 &  9.28E-06 & 117 &  35 \\
 2000.17 &  8.15E-06 &  5.25E-06 &  1.07E-05 &  43 &  31 \\
 2000.50 &  2.66E-05 &  1.15E-05 &  1.80E-05 & 122 &  42 \\
 2000.83 &  2.51E-05 &  9.22E-06 &  1.35E-05 & -31 &  24 \\
 2001.17 &  4.05E-05 &  1.60E-05 &  2.41E-05 &  15 &  19 \\
 2001.83 &  1.16E-05 &  9.58E-06 &  2.64E-05 & -82 &  35 \\
 2002.17 &  2.59E-05 &  1.12E-05 &  1.75E-05 &  56 &  18 \\
 2002.50 &  6.75E-05 &  1.72E-05 &  2.23E-05 &  34 &  17 \\
 2002.83 &  7.13E-05 &  2.33E-05 &  3.25E-05 &  14 &  25 \\
 2003.50 &  4.42E-05 &  2.11E-05 &  3.49E-05 &  -7 &  39 \\
 2003.83 &  6.59E-05 &  1.95E-05 &  2.64E-05 & -92 &  25 \\
 2004.17 &  3.73E-05 &  1.37E-05 &  2.01E-05 &  -6 &  30 \\
 2004.50 &  2.83E-05 &  1.54E-05 &  2.75E-05 &   3 &  32 \\
 2004.83 &  4.79E-05 &  1.49E-05 &  2.04E-05 & -45 &  15 \\
 2005.17 &  9.30E-05 &  2.17E-05 &  2.75E-05 &  29 &  16 \\
 2005.50 &  1.32E-04 &  2.73E-05 &  3.36E-05 &  48 &   8 \\
 2005.83 &  9.01E-05 &  1.90E-05 &  2.36E-05 &  41 &  13 \\
 2006.17 &  7.07E-05 &  1.70E-05 &  2.16E-05 &  19 &   7 \\
 2006.50 &  3.52E-05 &  1.29E-05 &  1.89E-05 &  47 &  18 \\
 2006.83 &  3.72E-06 &  3.08E-06 &  8.50E-06 &  -3 &  35 \\
 2007.17 &  1.00E-05 &  5.43E-06 &  9.71E-06 &  59 &  34 \\
 \hline
 \end{tabular}
 \end{table}

\begin{table}[h]
\caption{Same as Table~\ref{tab_3} but for the big particles with impact charges $ Q_I > 8.54 \times \mathrm{10^{-13}\,C}$. The data are shown in the bottom panel of Figure~\ref{fig_4} and in the bottom panel of Figure~\ref{fig_7}. 
 \label{tab_5} }
\vskip4mm
\centering
\tiny
\begin{tabular}{cccccc} \hline
  Time            &   Flux                     &  Err Flux-                    &   Err Flux+  & $\Delta \phi$  &   Err $\Delta \phi$ \\
$\mathrm{[Year]}$ &$\mathrm{[m^{-2}\,s^{-1}]}$ & $\mathrm{[m^{-2}\,s^{-1}]}$  &$\mathrm{[m^{-2}\,s^{-1}]}$ & $\mathrm{[deg]}$ & $\mathrm{[deg]}$ \\
  (1)   &    (2)              &   (3)   &    (4)    &    (5)    &    (6)   \\
 \hline
 1992.17 &  7.34E-05 &  1.94E-05 &  2.53E-05 &   1 &  12 \\
 1992.50 &  6.04E-05 &  1.72E-05 &  2.29E-05 &  17 &  11 \\
 1992.83 &  2.34E-05 &  1.01E-05 &  1.58E-05 &  10 &  21 \\
 1993.17 &  1.75E-05 &  8.35E-06 &  1.38E-05 &  -1 &  17 \\
 1993.50 &  4.40E-05 &  1.44E-05 &  2.01E-05 & -23 &  21 \\
 1993.83 &  4.79E-05 &  1.36E-05 &  1.82E-05 &  -1 &  12 \\
 1994.17 &  4.72E-05 &  1.34E-05 &  1.79E-05 & -18 &   9 \\
 1994.50 &  3.21E-05 &  1.27E-05 &  1.92E-05 & -22 &  25 \\
 1994.83 &  5.41E-05 &  1.43E-05 &  1.86E-05 &   6 &  12 \\
 1995.50 &  1.81E-05 &  9.80E-06 &  1.75E-05 &   3 &  12 \\
 1995.83 &  6.27E-05 &  2.31E-05 &  3.37E-05 &  21 &  12 \\
 1996.17 &  5.21E-05 &  1.54E-05 &  2.09E-05 &  16 &   9 \\
 1996.50 &  2.69E-05 &  1.07E-05 &  1.61E-05 &  17 &  10 \\
 1996.83 &  2.70E-05 &  1.16E-05 &  1.82E-05 &   3 &   7 \\
 1997.17 &  1.40E-05 &  7.58E-06 &  1.35E-05 & -41 &  37 \\
 1997.50 &  2.70E-05 &  1.07E-05 &  1.61E-05 &  -5 &   6 \\
 1997.83 &  2.32E-05 &  1.00E-05 &  1.57E-05 &  14 &  10 \\
 1998.17 &  2.26E-05 &  9.73E-06 &  1.52E-05 & -12 &  10 \\
 1998.50 &  2.31E-05 &  9.95E-06 &  1.56E-05 & -10 &  13 \\
 1998.83 &  1.29E-05 &  7.00E-06 &  1.25E-05 & -15 &   8 \\
 1999.17 &  1.76E-05 &  8.41E-06 &  1.39E-05 &   3 &  19 \\
 1999.50 &  1.45E-05 &  7.87E-06 &  1.41E-05 &  -5 &  27 \\
 1999.83 &  4.06E-06 &  3.36E-06 &  9.28E-06 &  -5 &  35 \\
 2000.17 &  2.04E-05 &  8.78E-06 &  1.38E-05 & -31 &   9 \\
 2000.50 &  1.06E-05 &  6.86E-06 &  1.40E-05 &  23 &  11 \\
 2000.83 &  4.65E-05 &  1.27E-05 &  1.68E-05 & -12 &  10 \\
 2001.17 &  3.37E-05 &  1.45E-05 &  2.28E-05 &   3 &  10 \\
 2001.83 &  3.47E-05 &  1.88E-05 &  3.36E-05 &  22 &   8 \\
 2002.17 &  3.11E-05 &  1.23E-05 &  1.85E-05 &  -9 &  13 \\
 2002.50 &  4.05E-05 &  1.32E-05 &  1.85E-05 &   0 &   9 \\
 2002.83 &  3.17E-05 &  1.51E-05 &  2.50E-05 &   9 &  14 \\
 2003.50 &  2.21E-05 &  1.42E-05 &  2.90E-05 &  38 &  12 \\
 2003.83 &  5.99E-06 &  4.96E-06 &  1.37E-05 &  18 &  35 \\
 2004.17 &  1.60E-05 &  8.67E-06 &  1.55E-05 & -18 &  26 \\
 2004.50 &  9.45E-06 &  7.83E-06 &  2.16E-05 & -82 &  35 \\
 2004.83 &  2.87E-05 &  1.14E-05 &  1.71E-05 &   2 &  14 \\
 2005.17 &  5.17E-06 &  4.28E-06 &  1.18E-05 &   8 &  35 \\
 2005.50 &  3.44E-05 &  1.36E-05 &  2.05E-05 &  55 &  19 \\
 2005.83 &  6.14E-05 &  1.57E-05 &  2.03E-05 &  29 &  10 \\
 2006.17 &  5.41E-05 &  1.48E-05 &  1.95E-05 &  14 &   8 \\
 2006.50 &  2.51E-05 &  1.08E-05 &  1.70E-05 &  17 &   6 \\
 2006.83 &  7.43E-06 &  4.79E-06 &  9.76E-06 &   2 &  13 \\
 2007.17 &  1.00E-05 &  5.43E-06 &  9.71E-06 &   5 &  11 \\
 \hline
 \end{tabular}
 \end{table}

\clearpage

\section*{Figures}

\begin{figure}
   \centering
\parbox{15cm}{
\vspace{-5cm}
\hspace{-2cm}
\includegraphics[width=1.1\textwidth]{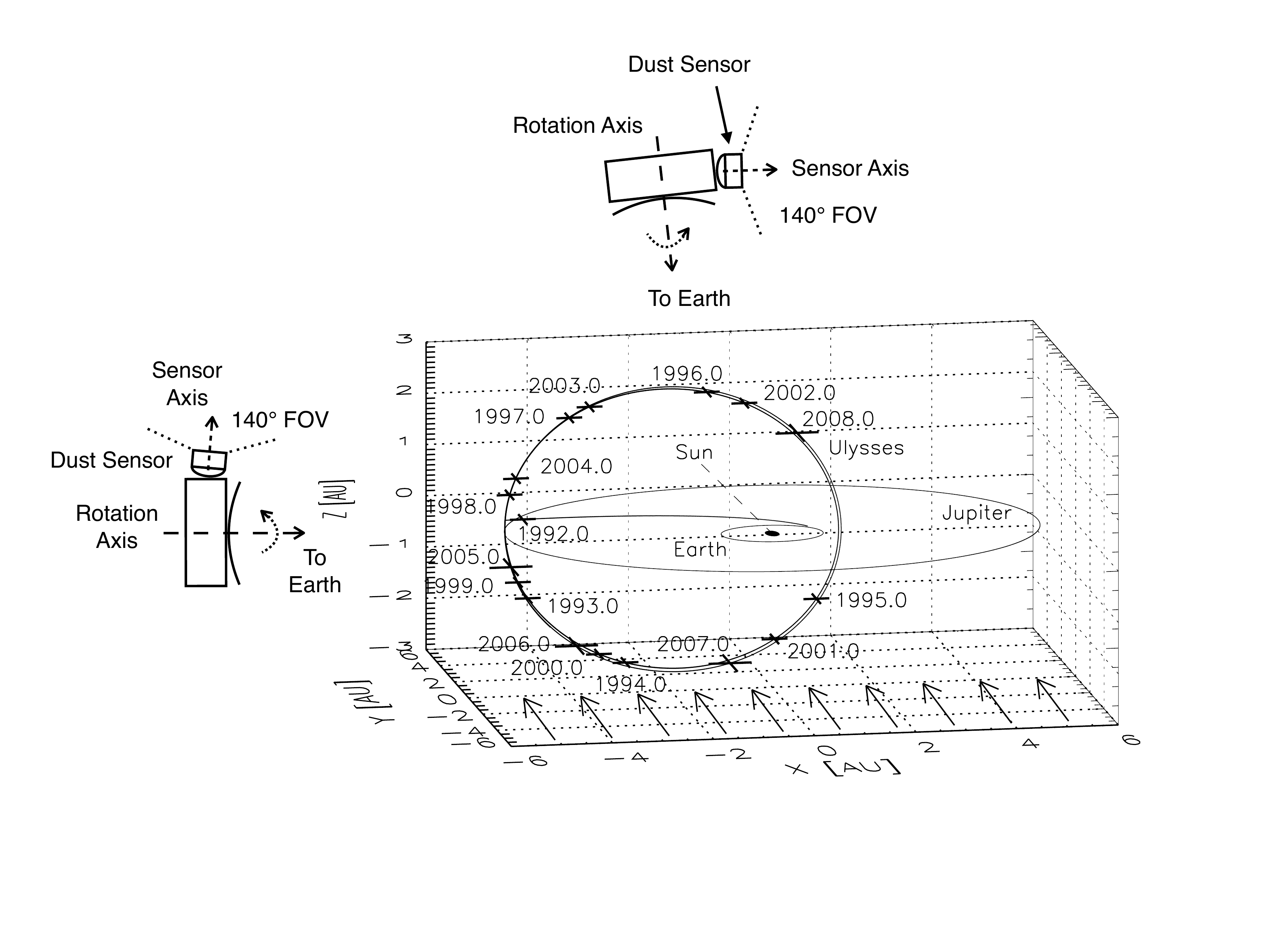}
}
\caption[]{
The trajectory of Ulysses in ecliptic coordinates with the Sun at the center. The orbits of Earth and Jupiter indicate the ecliptic plane, and the initial trajectory of Ulysses was in this plane. After Jupiter flyby in early 1992 the orbit was almost perpendicular to the ecliptic plane ($79^{\circ}$ inclination). Crosses mark the spacecraft position at the beginning of each year. Vernal equinox is to the right (positive X axis). Arrows indicate the undisturbed interstellar dust flow direction which is within the measurement accuracy co-aligned with the direction of the interstellar helium gas flow. It is almost perpendicular to the orbital plane of Ulysses. The orientation of the spacecraft's rotation axis and the dust instrument bore-sight are indicated for both an orbital position near the aphelion and above the solar north pole.
}
\label{fig_3}
\end{figure}

\begin{figure}
   \centering
\parbox{15cm}{
\vspace{-1cm}
\includegraphics[width=0.8\textwidth]{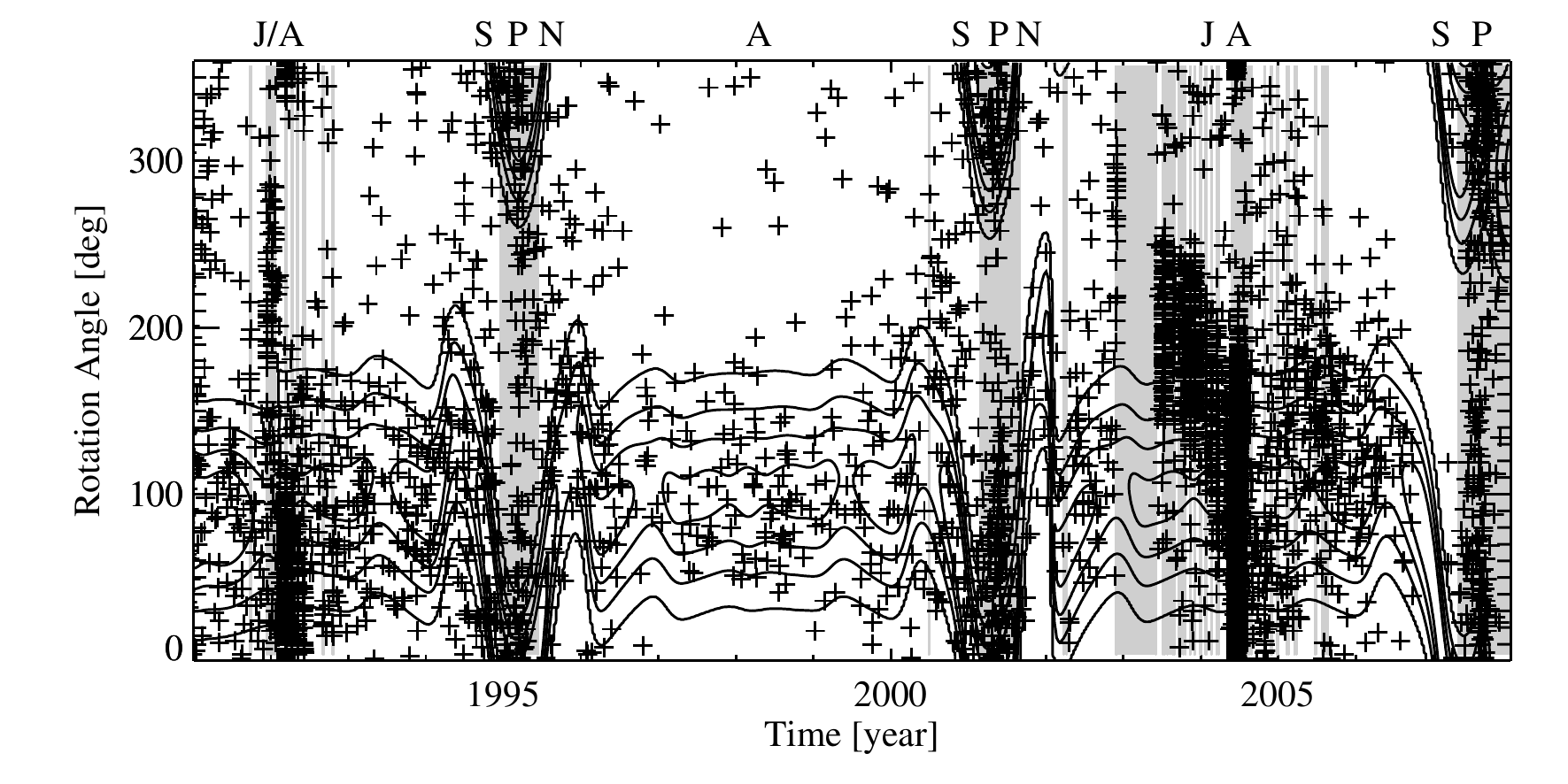}
}
\parbox{15cm}{
\includegraphics[width=0.8\textwidth]{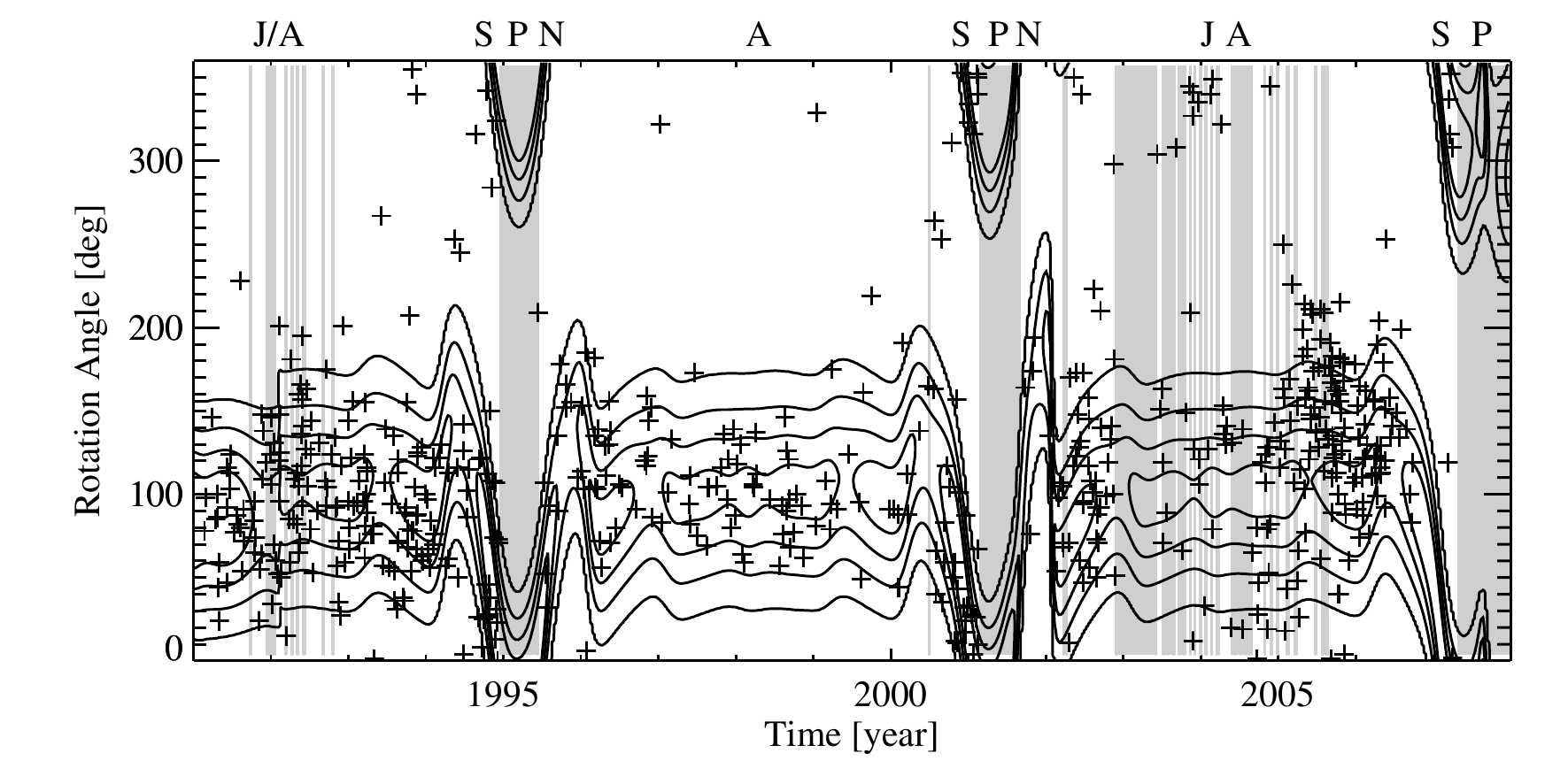}
}
\caption[]{
Impact direction of the dust particles detected by Ulysses over time. The upper panel shows all dust impacts detected during the Ulysses mission. The lower panel shows the particles identified as ISD. Each cross corresponds to a dust particle impact. The time intervals highlighted with grey background correspond to periods when no ISD data is available due to filtering of the perihelion passages (1995, 2001), several instrument switch-offs \citep{krueger2006b,krueger2010b}, or Jovian dust streams. In 1998, when Ulysses was close to Jupiter's orbit, no dust streams were observed because Jupiter was on the opposite side of its orbit. The contour lines in both panels show the dust detector's sensitivity for particles arriving from the nominal (undeflected) ISD direction. Labels at the top indicate Ulysses' Jupiter flybys (J), perihelion passages (P), aphelion passages (A), south polar passes (S) and north polar passes of Ulysses (N). {\em Top panel:} The entire Ulysses dust data set.  The periods of high dust impact rates in 1992 and 2003 to 2004 are due to Jovian dust streams \citep{krueger2006c}. {\em Bottom panel:} The dust impacts identified as interstellar particles. The selection criteria are described in the text and summarised in Table~\ref{tab_1}.
}
\label{fig_1}
\end{figure}

\begin{figure}
   \centering
\parbox{15cm}{
\includegraphics[width=0.92\textwidth]{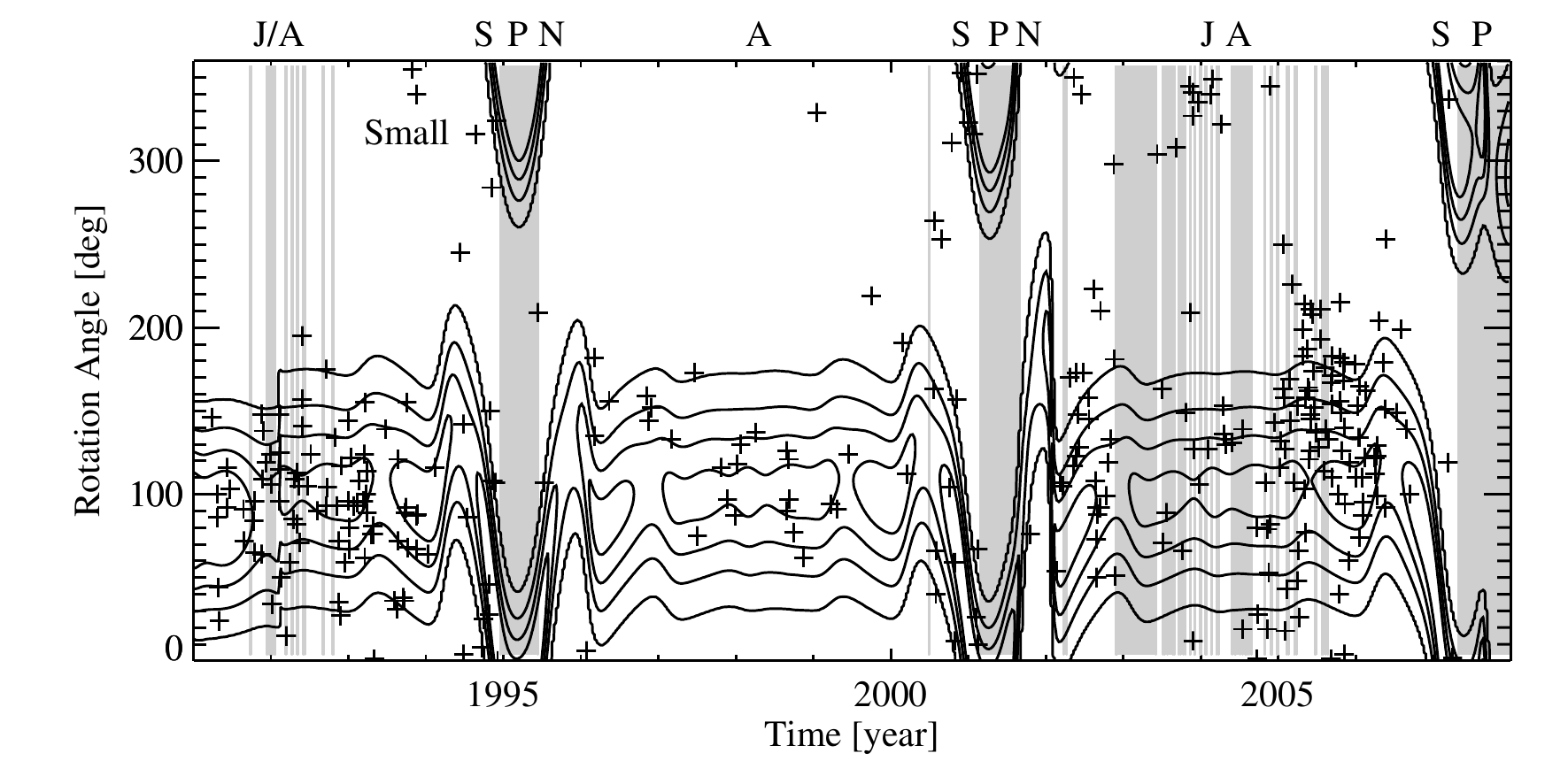}
}
\parbox{15cm}{
\includegraphics[width=0.92\textwidth]{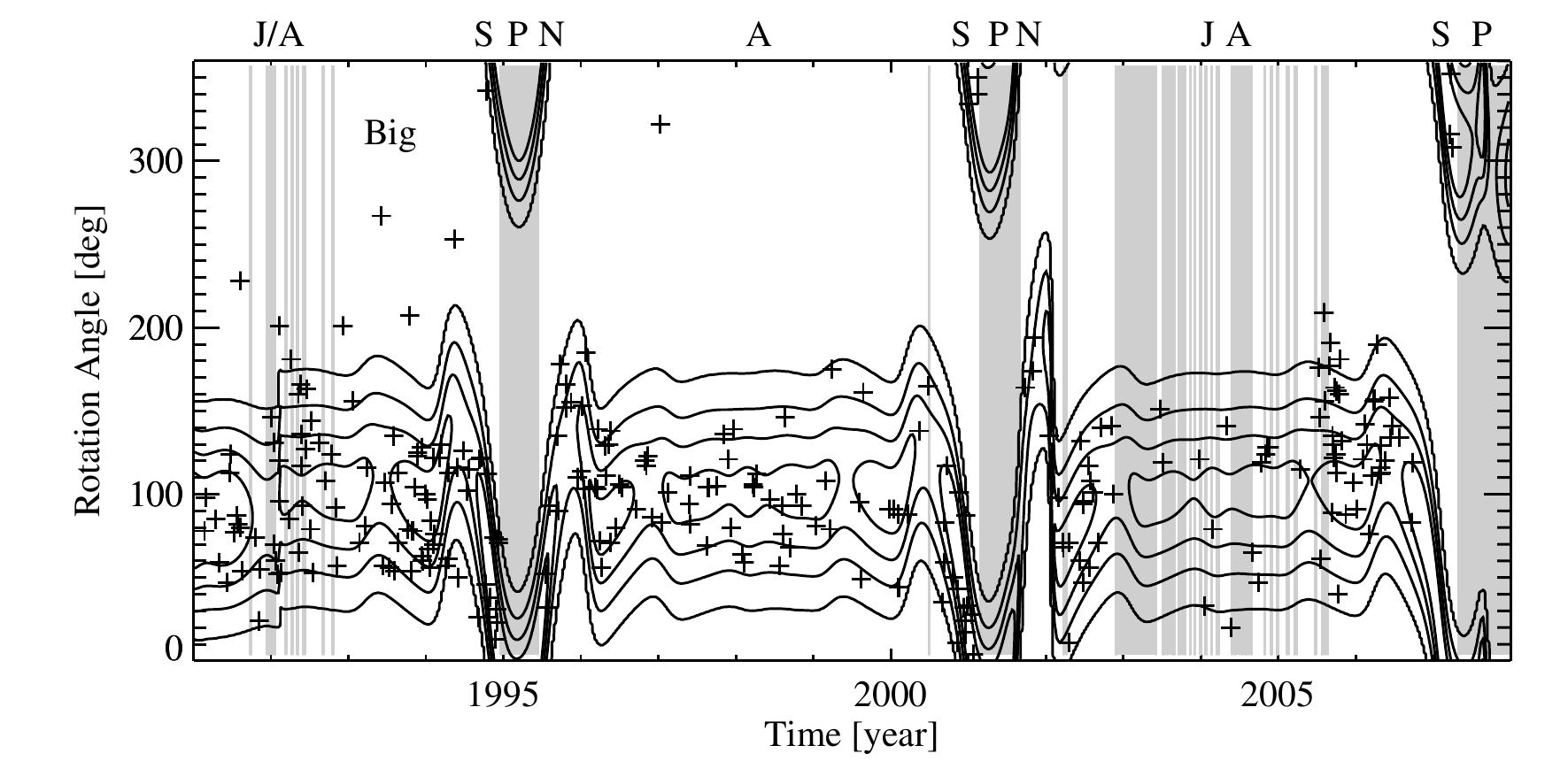}
}
\caption[]{
Same as Figure~\ref{fig_2} for two subsets of the Ulysses ISD data containing small ({\em top panel}, 273 particles, $\mathrm{10^{-13}\,C} <  Q_I \leq  8.54 \times 10^{-13}\,\mathrm{C}$)  and large particls ({\em bottom panel}, 307 particles, $Q_I > 8.54\,\times 10^{-13}\,\mathrm{C}$), respectively. The selection criteria for small and large particles are described in Section~\ref{sec_preparation}.
}
\label{fig_2}
\end{figure}

\begin{figure}
   \centering
\parbox{15cm}{
\includegraphics[width=0.95\textwidth]{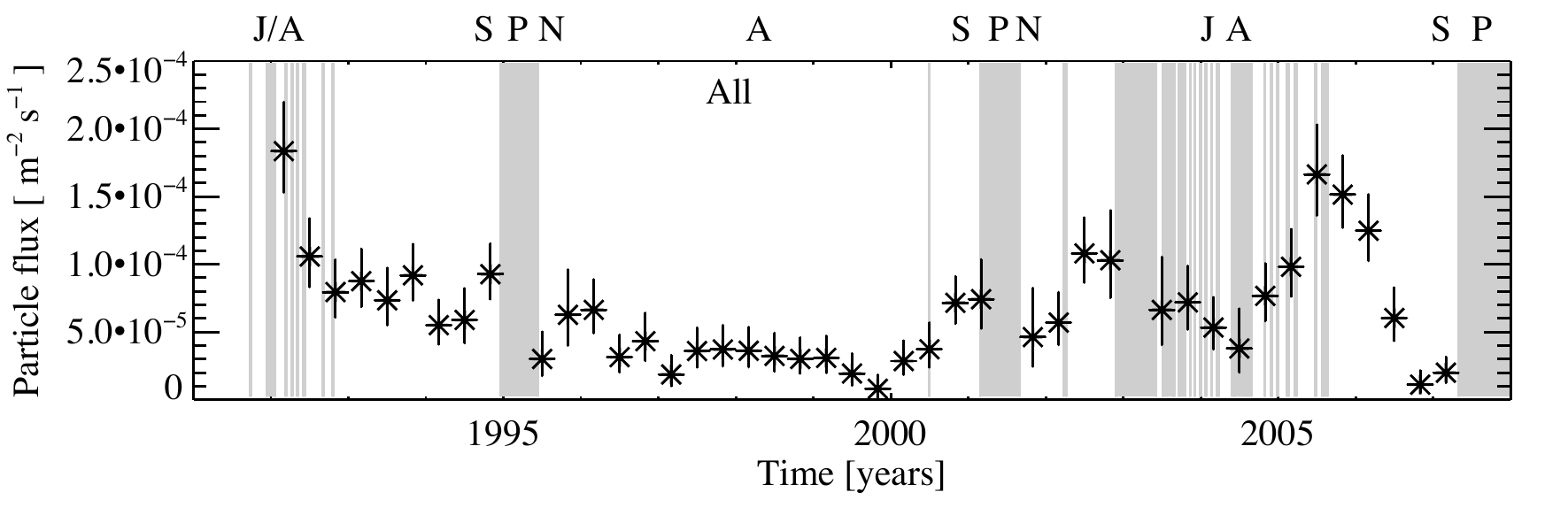}
}
\parbox{15cm}{
\includegraphics[width=0.95\textwidth]{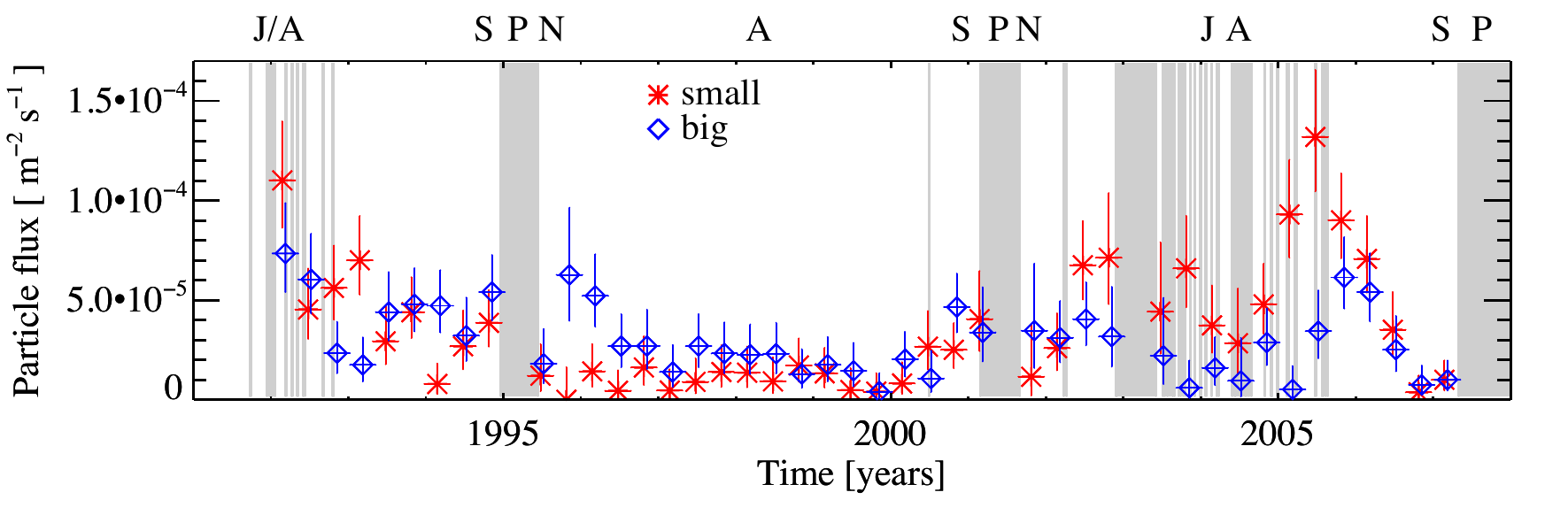}
}
\caption[]{
Measured flux of the ISD particles. {\em Top panel:} Flux of all ISD particles. {\em Bottom panel:} Flux of the small and large particles (divided as described in Section~\ref{sec_preparation}). Red asterisks correspond to the small particles, blue diamonds show the large particles.
}
\label{fig_4}
\end{figure}

\begin{figure}
   \centering
\parbox{15cm}{
\includegraphics[width=0.95\textwidth]{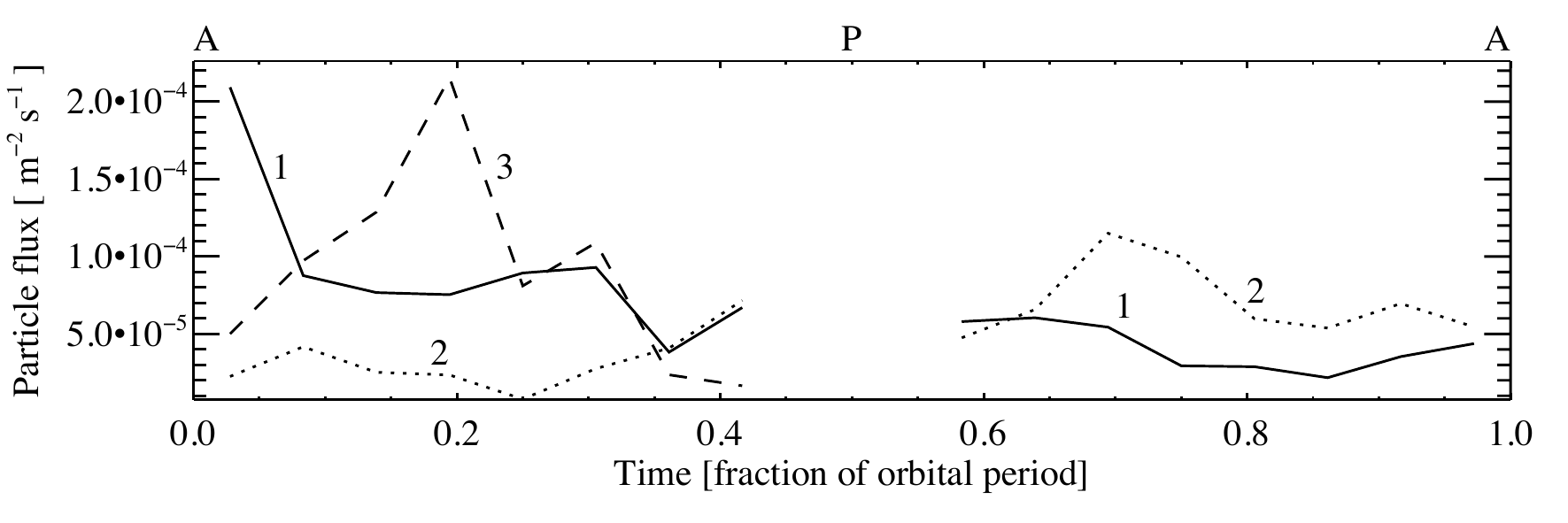}
}
\caption[]{
ISD flux per orbit. 
Aphelion is at 0.0 and 1.0, perihelion is at 0.5. The solid line shows the orbit from 1992 to 1998 (orbit 1), the dotted line the orbit from 1998 to 2004 (orbit 2), and the dashed line denotes the orbit from 2004 to 2010 (orbit 3). Given that the dust detector was switched off permanently in 2007, no dust data were measured during the second half of the last orbit.
}
\label{fig_5}
\end{figure}

\begin{figure}
   \centering
\parbox{15cm}{
\includegraphics[width=0.95\textwidth]{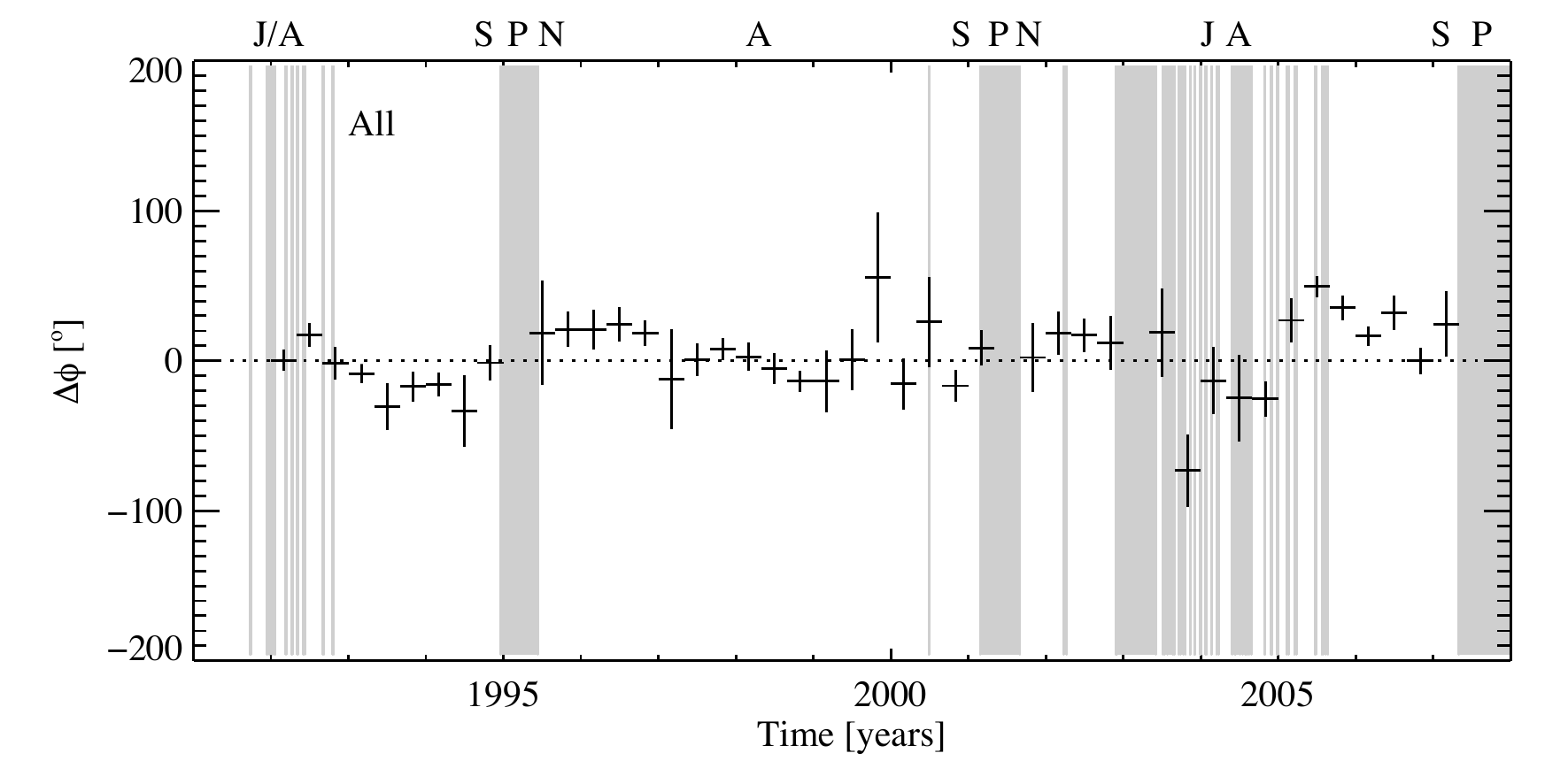}
}
\caption[]{
Deviation of the measured dust impact direction from the flow direction of interstellar helium. The deviation is given in the spacecraft rotation angle coordinate, as described in the text. The interstellar helium direction \citep[$0^{\circ}$ in this plot;][]{witte2004a} is in good agreement with the average direction of the dust flow. 
}
\label{fig_6}
\end{figure}

\begin{figure}
   \centering
\parbox{15cm}{
\includegraphics[width=0.95\textwidth]{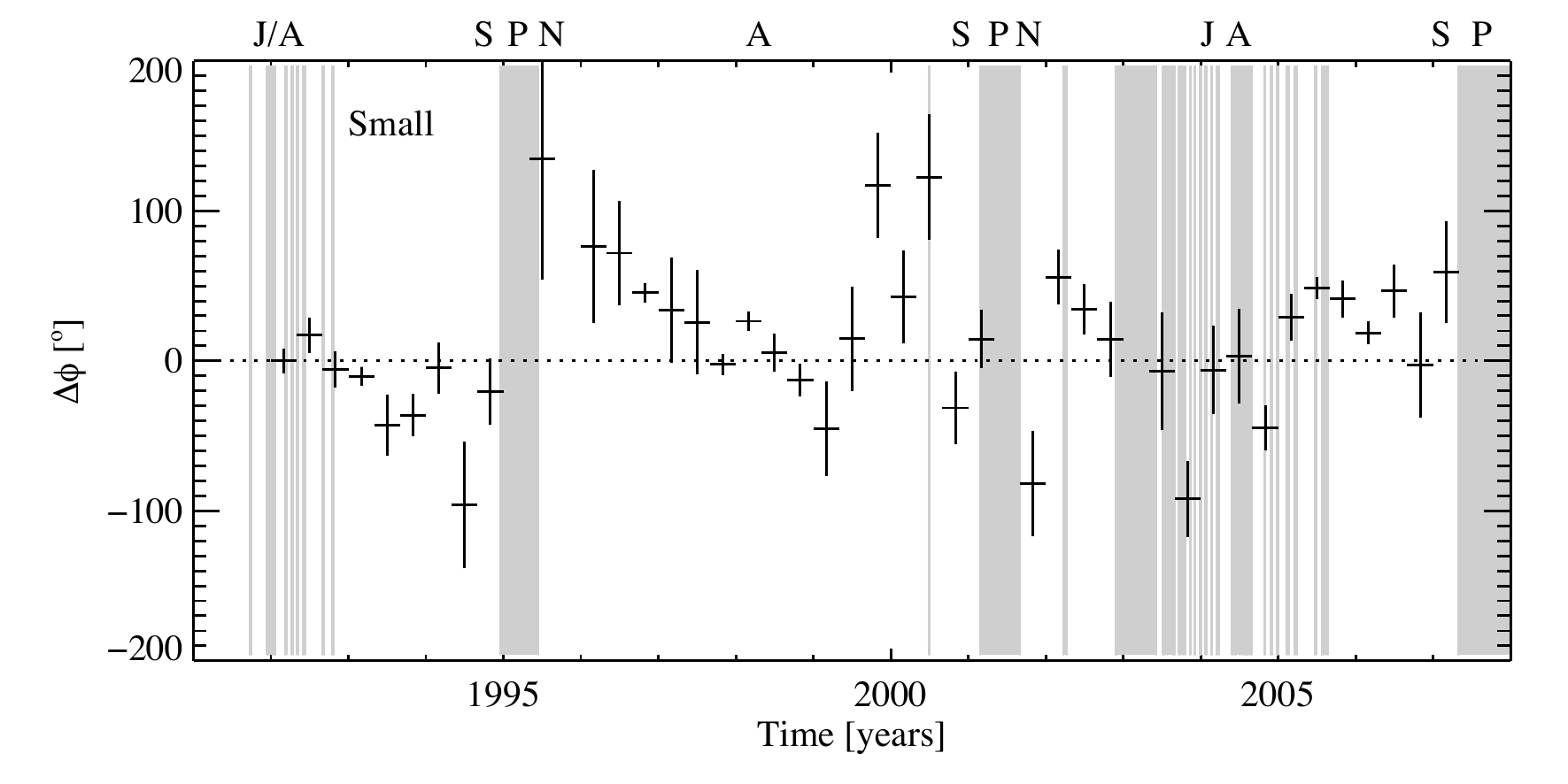}
}
\parbox{15cm}{
\includegraphics[width=0.95\textwidth]{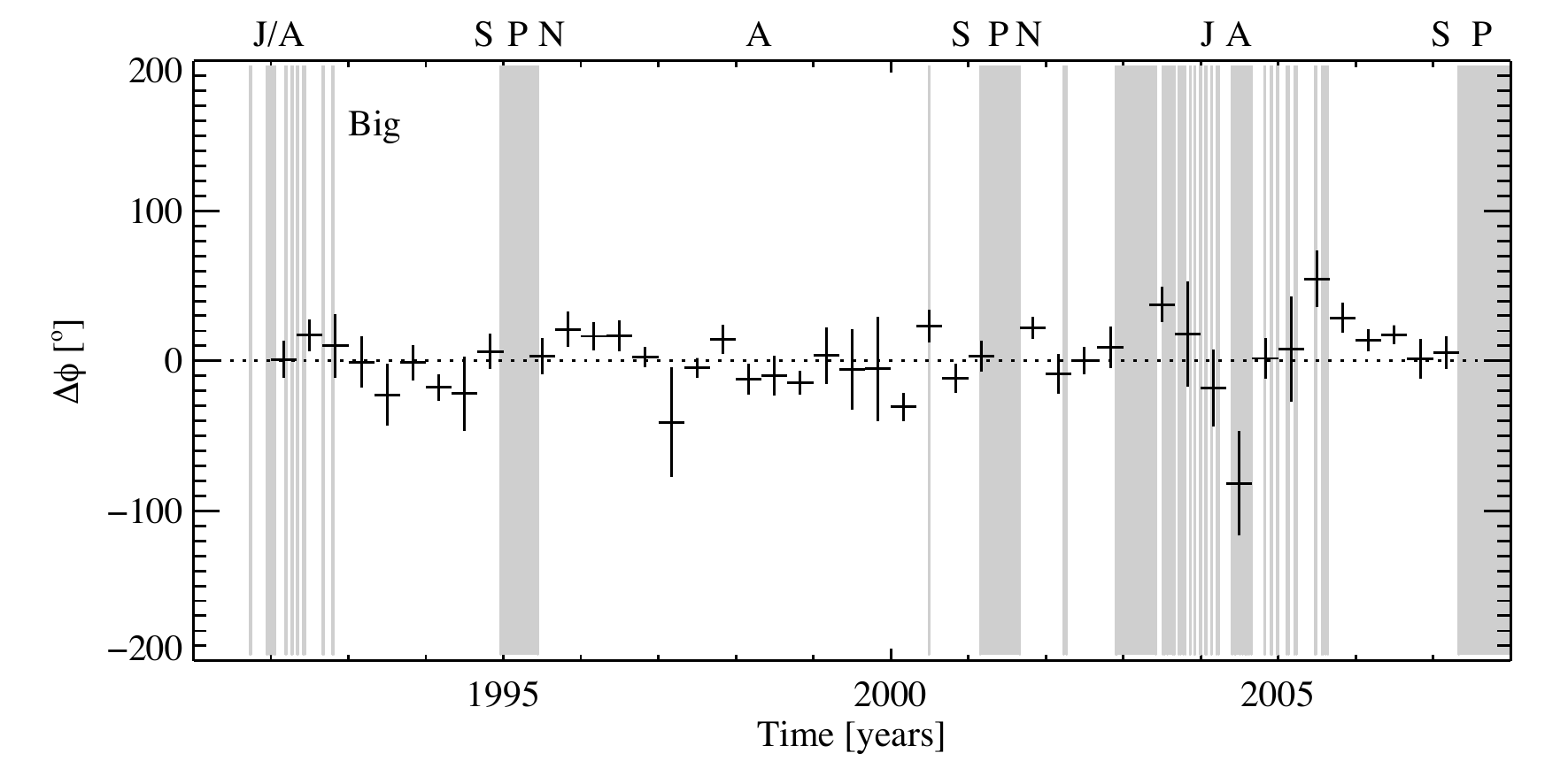}
}
\caption[]{
Same as Figure~\ref{fig_6}, but separately for small particles ({\em upper panel}) and  large particles ({\em lower panel}). The definition of small and large particles is described in Section~\ref{sec_preparation}.
}
\label{fig_7}
\end{figure}

\begin{figure}
   \centering
\includegraphics[width=0.9\textwidth]{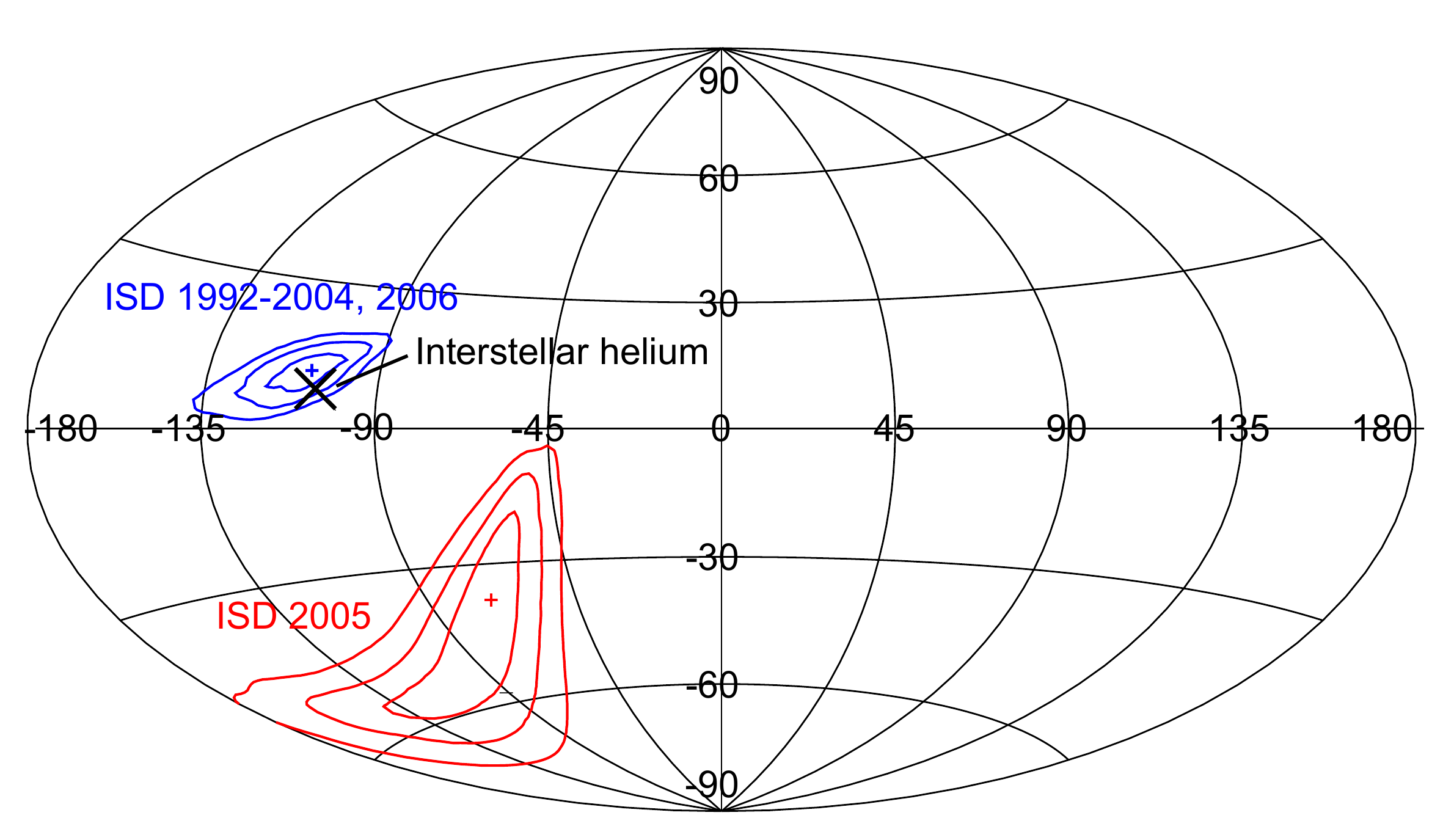}
\caption[]{
Flow direction of the interstellar dust in ecliptic coordinates. Data from 2005 when the dust was deflected away from the interstellar helium direction are shown in red, data from the rest of the mission when no significant deflection occurred (1992 to 2004 and 2006) are shown in blue. The three contour lines correspond to confidence levels of $1\sigma$, $2 \sigma$, and $3 \sigma$, respectively. The large black cross indicates the flow direction of the interstellar helium through the solar system, which is in good agreement with the overall dust flow direction.
}
\label{fig_8}
\end{figure}

\begin{figure}
   \centering
\parbox{15cm}{
\includegraphics[width=0.95\textwidth]{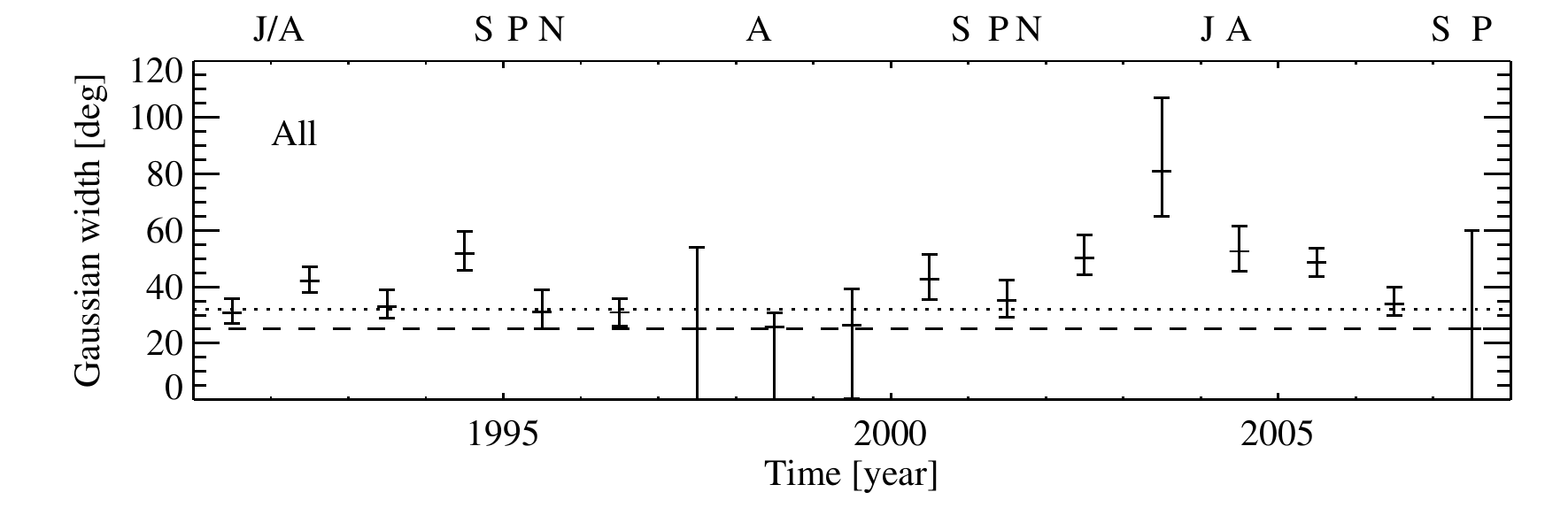}
}
\parbox{15cm}{
\includegraphics[width=0.95\textwidth]{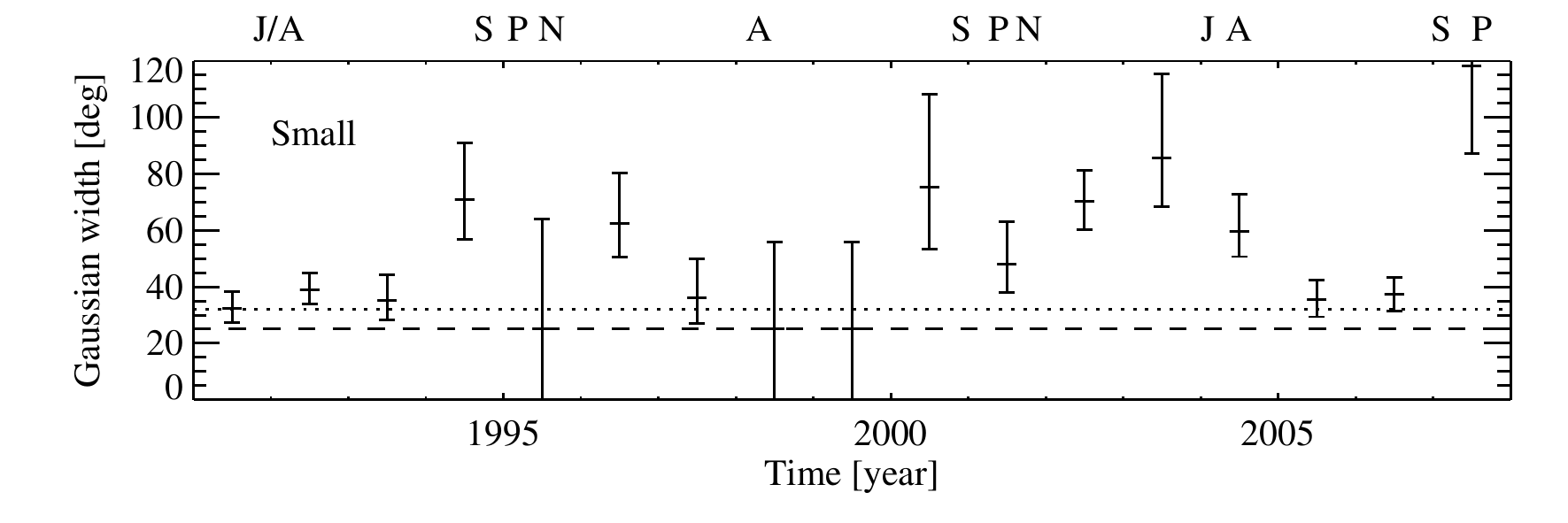}
}
\parbox{15cm}{
\includegraphics[width=0.95\textwidth]{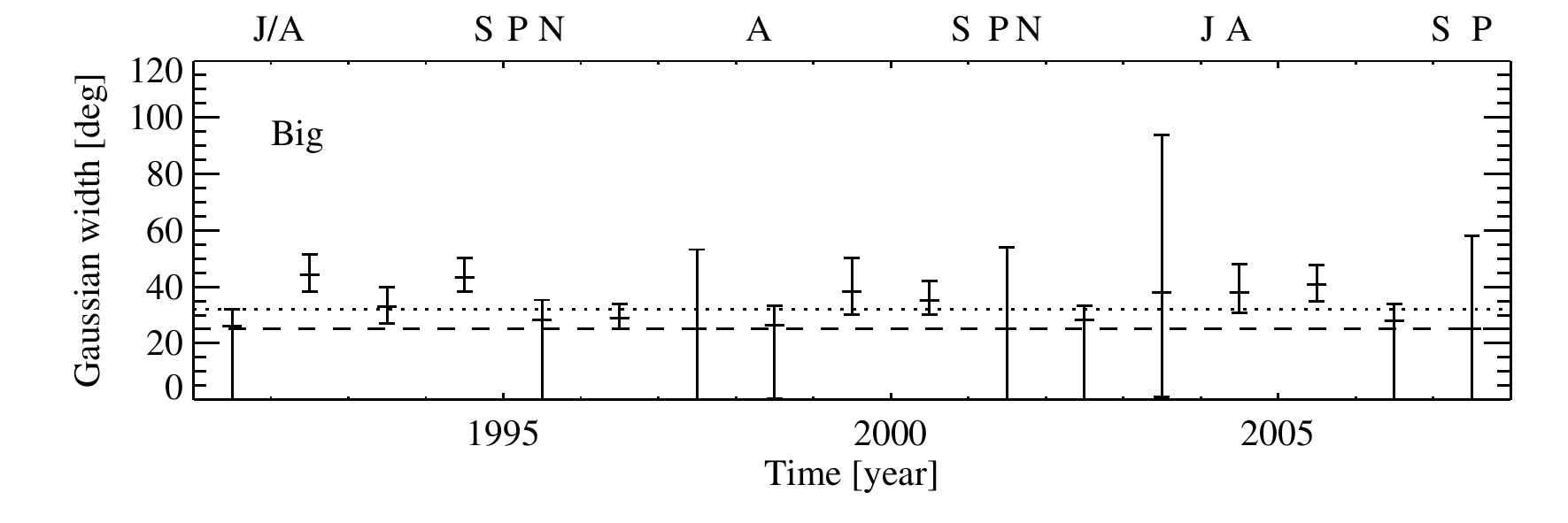}
}
\caption[]{
Gaussian width of the ISD flow in 1~yr intervals. The {\em upper panel} shows the width for all particles, the {\em middle panel} that for small particles, and the {\em bottom panel} shows the width for large particles. The dashed line shows the expected width of a monodirectional stream broadened by the detector's angular sensitivity profile (sensor target only), and the dotted line indicates the detector's sensitivity profile including wall impacts. The error bars indicate the $1 \sigma$-uncertainties of the width of the fitted Gaussian. The determination of the angular width  is described in Section~\ref{sec_width}.
}
\label{fig_9}
\end{figure}

\begin{figure}
   \centering
\parbox{15cm}{
\includegraphics[width=0.95\textwidth]{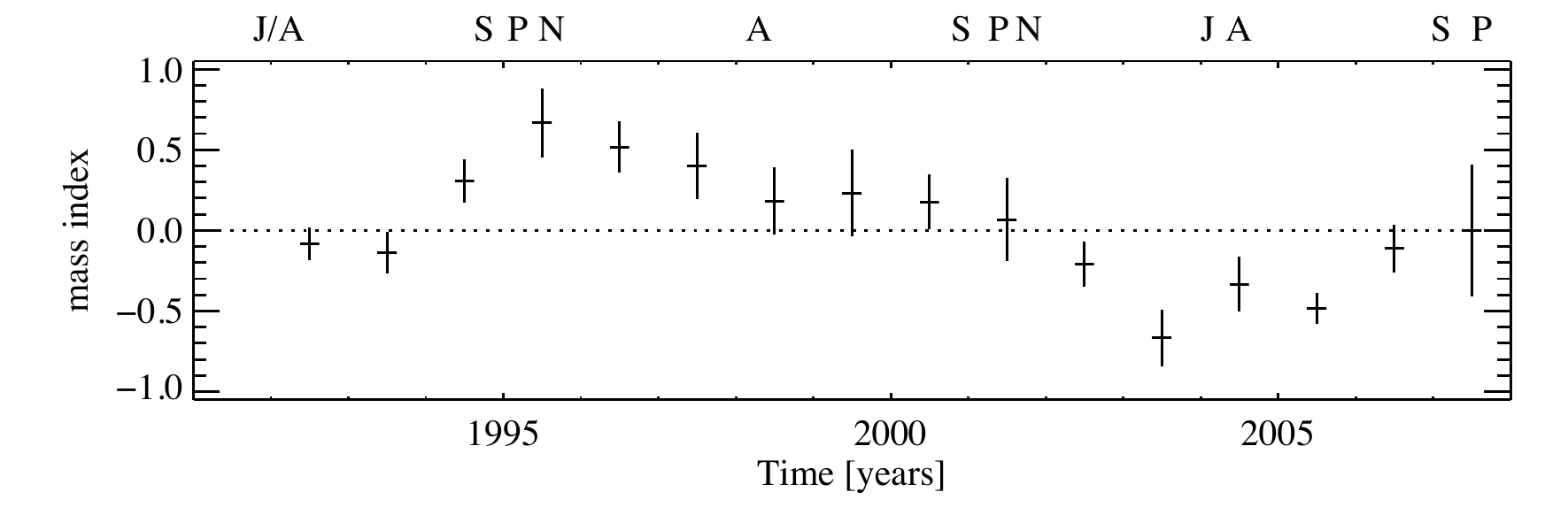}
}
\caption[]{
Mass index of the ISD population over time for intervals of 1~yr. The mass index MI is defined in Eq.~\ref{equ_4}. The error bars indicate $1\sigma$ uncertainties determined from error propagation of the Poissonian errors. The separation into small and large particles is based on the measured ion impact charge as described in Section~\ref{sec_preparation}. The solar cycle is indicated at the top.
}
\label{fig_10}
\end{figure}

\end{document}